\documentclass[preprint,amsmath,amssymb,aps,pre,longbibliography]{revtex4-1}

\usepackage{graphicx}
\usepackage{dcolumn}
\usepackage{bm}
\usepackage{color}
\usepackage[utf8]{inputenc}
\usepackage{cancel}

\begin{document}

\title{Viscosity as the product of its ideal low-concentration value times a thermodynamic function}

\author{L. Marchioni}
\author{M. A. Di Muro}  
\author{M. Hoyuelos}

\email{hoyuelos@mdp.edu.ar}

\affiliation{Instituto de Investigaciones Físicas de Mar del Plata (IFIMAR -- CONICET), Departamento de Física, Facultad de Ciencias Exactas y Naturales,
	Universidad Nacional de Mar del Plata, Deán Funes 3350, 7600 Mar del Plata, Argentina}

\date{\today}

\begin{abstract}
The behavior of viscosity, $\eta$, as a function of concentration in dense fluids remains an unsolved problem, as is the case with other transport coefficients. Boltzmann's theory and the Chapman-Enskog method predict the value of the viscosity at low concentrations, $\eta_0$. Here, the hypothesis $\eta=\phi\, \eta_0$ is proposed, where $\phi$ is a function of the thermodynamic state that represents the effects of interactions as concentration increases. We consider that $\eta_0$ is the viscosity in an ideal hypothetical system, where the condition of small interactions applies for the whole density range ($\phi \to 1$ for low concentration). The method proposed to verify this hypothesis involves coupling the system with a solvent represented by a Langevin thermostat, characterized by a damping time $t_d$. Molecular dynamics simulations show that different values of noise intensity modify $\eta$ and $\eta_0$, but do not affect $\phi$. This result supports the assumption that $\phi$ is a state function, since the thermodynamic state remains unaltered by the presence of damping and noise. Simulations were conducted for particles that interact via a pseudo-hard sphere or a Lennard-Jones potential.
\end{abstract}


\maketitle

\section{Introduction}

The study and understanding of transport phenomena are highly relevant to general research and to many areas of industry.
Transport features of fluids can be categorized into three main types. Diffusivity relates to the displacement of particles or molecules throughout the fluid, typically composed of particles of the same type. 
Thermal conductivity accounts for the exchange of kinetic energy between the constituents of the fluid that arises from collisions. 
Finally, viscosity quantifies the friction that occurs between two adjacent layers of a fluid where the molecules move on average with different velocities. More precisely, the viscosity coefficient indicates the momentum transfer rate between layers. 
For dilute gases, closed expressions can be derived using the Chapman-Enskog theory \cite{chapman}, which is based on Boltzmann's equation. An analytical description covering the entire density range has proven to be a formidable challenge. 
There have been several attempts to address this issue, particularly for viscosity and other transport coefficients. 
Enskog extended his theory to dense fluids of hard spheres \cite{enskog}, where the viscosity depends on the radial distribution function $g(\sigma)$ and the second virial coefficient $b$; $\sigma$ represents the diameter of the spheres. A dense fluid is said to be a system of particles with a sufficiently high density that intermolecular interactions become significant, which makes the behavior of the system to deviate from the ideal gas assumptions. However, Enskog's theory is only valid for small to moderate densities because one of the assumptions in the formulation is the molecular chaos approximation. 

Koo and Hess derived an expression for dense hard sphere fluids using the Kirkwood-Smoluchowski equation \cite{Koo}, which agrees with simulation results but fails at low densities.
Other authors have attempted to adapt the theory of Enskog to make it suitable for real fluids \cite{hanley}; the result is known as \textit{Modified Enskog Theory} (MET). It assumes that a real fluid generally exhibits a behavior similar to that of a hard sphere fluid, specially for small or moderate concentrations. The values of $g(\sigma)$ and $b$ are obtained from existing equations of state for real fluids. 
In this context, the viscosity of hard spheres, $\eta_{HS}$, is related to the Enksog viscosity, $\eta_{E}$, by the expression $\eta_{HS}=F(\rho^*)\eta_E$, where $\rho^*$ is the reduced density and $F(\rho^*)$ is a correction function. Additionally, the HS viscosity can be expressed as $\eta_{HS}=G(\rho^*)\eta_0$, where $\eta_0$ is the viscosity at low concentration and $G(\rho^*)=F(\rho^*)\eta_E/\eta_0$. The dense gas and liquid regions have been analyzed by Dymond \cite{dymond,dymond2,dymond3,Dymond1996}. Different expressions for the correction function have been proposed by Silva \cite{SilvaCoelho} and by Heyes and Sigurgeirsson \cite{heyes3}.
A different proposal to improve Enskog's theory proposed by Beijeren and Ernst is known as Revised Enskog theory (RET)\cite{Beijeren2}, in which a spatially correlated radial distribution function is used, so as to capture the properties of nonuniform systems. This approach has been used to obtain the transport coefficients for multicomponent mixtures \cite{lopezharo84} and hard sphere crystals \cite{Kirkpatrick}. Recently, Pieprzyk \textit{et al.} have evaluated the transport coefficients of a hard sphere fluid and crystal by intensive Molecular Dynamics simulations \cite{Pieprzyk2024}; they found that there is a good agreement with the results predicted by the RET.

The effective Hard-Sphere Diameter method represents a real fluid as a purely repulsive hard sphere system and then considers the attractive forces as a perturbation. For instance, the Weeks-Chandler-Andersen perturbation theory (WCA) \cite{weeks} splits the Lennard-Jones potential into a reference term that contains only the repulsive forces, and a perturbation term that accounts for the attractive forces that manifest in a real fluid. Silva \textit{et al.} provided an approximate expression for the viscosity of a fluid with WCA interactions \cite{SilvaCoelho}  

Another approach to computing viscosity and transport coefficients in general consists of free volume theories; they provide relatively simple expressions that depend on a quantity called free volume, $V_f=V-V_i$, where $V$ is the volume and $V_i$ is an intrinsic molar volume. Free volume theories are generally valid only for systems with repulsive interactions, although they can be extended to include attractive interactions if the concept of activation energy is considered. Several models have been proposed, including those by Dymond-Hildebrand-Batschinski \cite{dymond,hildebrand,batschinski}, Doolittle–Cohen–Turnbull \cite{doolittle,cohen2}, Turnbull and Cohen \cite{turnbull}, and Macedo-Litovitz \cite{macedo}.

A different method that yields good results involves excess entropy scaling laws. Rosenfeld \cite{rosenfeld,rosenfeld2,dyre} derived a relation between a dimensionless viscosity $\tilde{\eta}$ and the excess entropy $S^{\text{ex}}$: $\tilde{\eta}=0.2\,\text{exp}(-0.8 \,S^{\text{ex}}/N k_B)$, where $N$ is the number of particles and $k_B$ is the Boltzmann constant (more details on how to calculate reduced dimensionless quantities are provided below). This relation has proven to be quasi-universal, showing good agreement with simulations for various types of pair potentials, particularly in the moderate and large density regimes.

Even though we have delved into different approaches for calculating viscosity, it is worth mentioning that these methods can also be applied to obtain approximate expressions for self-diffusivity and thermal conductivity.

As we see, a significant effort has been put to provide insight into the transport process theory. So far however, finding general expressions for the transport coefficients valid for the whole density range has proven to be an arduous task. In this manuscript we aim to better understand the role that interactions play in describing viscosity for dense systems. In Ref.~\cite{marchioni} it was suggested that the self-diffusion coefficient $D$ may be written as $D=\varphi D_0$, where $D_0$ is the self-diffusion coefficient at low densities (which can be calculated via the kinetic theory), and $\varphi$ is a function that depends only on the thermodynamic state of the system. This implies that microscopic parameters, such as the mass or size of the particles, do not explicity appear in general expressions of $\varphi$; instead, $\varphi$ encapsulates the \textit{macroscopic} information of the interactions. Here, we seek to verify whether an equivalent relation for the shear viscosity also holds: 
\begin{equation}\label{e.etaphi}
\eta=\phi\, \eta_0,
\end{equation}
where $\eta_0$ is the low-density viscosity and $\phi$ is an unknown function of the thermodynamic state.

The procedure is analogous to that described in \cite{marchioni}. We assume a system of interacting particles subjected to a Langevin thermostat. That is, the particles are immersed in a solvent composed of smaller ones, which apply both a stochastic force and a damping force. The idea is that the presence of noise should not have an effect on function $\phi$, since we assume that it is a thermodynamic function, and noise does not modify the thermodynamic state. Conversely, noise intensity is expected to affect the low concentration viscosity $\eta_0$, as it depends on the microscopic details of the model.

Cohen and de Schepper \cite{cohen3}, and also Heyes and Sigurgeirsson \cite{heyes3}, analyzed an analogy between a dense hard-sphere fluid and a colloidal suspension that uses as a key ingredient a relationship between the shear viscosity and self-diffusion. In a simulation, the solvent viscosity of the colloidal suspension can be progressively decreased giving place to a transition from Langevin to Newtonian dynamics \cite{heyes3}. An equivalent procedure is proposed here where, instead of decreasing the viscosity of the solvent, we consider a progressive increase of the damping time (Stokes formula indicates that the solvent viscosity is inversely proportional to the damping time). Nevertheless, there are important differences with our model. We neglect hydrodynamic forces (usually relevant in colloidal suspensions) and we do not assume the existence of a formal relationship between the shear viscosity and self-diffusion. The model that we use is a particle system (with hard sphere or Lennard-Jones interactions) with a Langevin thermostat; the feature that is interesting for us is that this model has an equation of state independent of the noise intensity, meaning that the thermodynamic state is not modified by the Langevin thermostat. The thermostat modifies the dynamics and the microscopic features, but not the thermodynamic state and, according to our hypothesis, not the function $\phi$.

We performed molecular dynamics simulations using LAMMPS software \cite{plimpton}, in which the velocity form of the Stoermer-Verlet time integration algorithm for constant particle number, volume, and energy (NVE ensemble) is used (more details in appendix \ref{Ap.1}).

The paper is organized as follows. In Sec.\ \ref{s.hyp} we present the motivation and justification for Eq.\ \eqref{e.etaphi} and describe the method used to assess its validity. In Sec.\ \ref{s.eta0}, viscosity at low concentration $\eta_0$ in the presence of noise is derived from the combination of Boltzmann and Langevin theories. In Sec.\ \ref{s.hs}, the ratio $\eta/\eta_0$ for hard spheres is numerically calculated as a function of density and damping time. In Sec.\ \ref{s.LJ}, we conduct a similar analysis for the Lennard-Jones potential at two different temperatures. Finally, conclusions are presented in Sec.\ \ref{s.conclusions}.

\section{The hypothesis and the method}
\label{s.hyp}

In Ref.\ \cite{marchioni}, it was proposed that the self-diffusion coefficient obeys a general equation of the form $D = \varphi D_0$; the proposal was motivated by earlier work on transition probabilities of tagged particles between two neighboring small cells \cite{dimuro}. It was demonstrated \cite{marchioni} that numerical simulations are consistent with the hypothesis that $\varphi$ is a state function, independent of microscopic details. Since this is an important guide for the development of a general transport theory for dense fluids, a natural step is to verify whether s similar equation holds for other transport coefficients. Here, we propose that viscosity conforms to a similar form, as given by Eq.\ \eqref{e.etaphi}, with $\phi$ a state function.

The method to test this hypothesis is as follows. We consider a system of $N$ particles in a volume $V$, interacting with each other via a given potential, with the temperature kept constant by a heat reservoir. Let $\mathbf{f}_i$ denote the force exerted on particle $i$ by the rest of the system. The particle system is immersed in a solvent, with  its effect represented by a Langevin thermostat. In this setup, the total force on particle $i$ is 
\begin{equation}\label{e.force}
\mathbf{F}_i = \mathbf{f}_i - m \mathbf{v}_i/t_d + \bm{\xi}_i
\end{equation}
where $t_d$ is the damping time, $m$ is the particle mass and $\mathbf{v}_i$ is the particle velocity; the term $- m \mathbf{v}_i/t_d$ represents friction and $\bm{\xi}_i$ is a stochastic force with zero mean ($\langle \xi_i(t) \rangle =0 $) and delta time correlated ($\langle \xi_i(t) \xi_i(t') \rangle = \frac{2 m k_B T}{t_d}\delta(t-t') $). If the temperature and the damping time are known, the noise intensity is determined by the fluctuation-dissipation theorem. The coupling to the thermostat is effectively represented by the damping time $t_d$.

\begin{figure}
	\begin{center}
	a)\includegraphics[width=7.5cm]{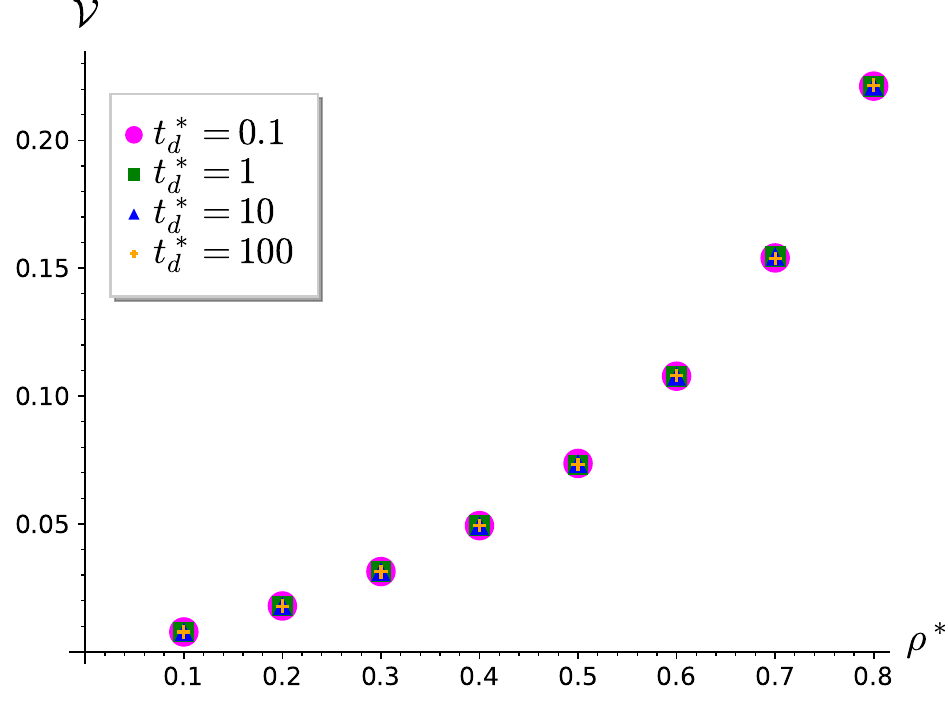}
        b)\includegraphics[width=7.5cm]{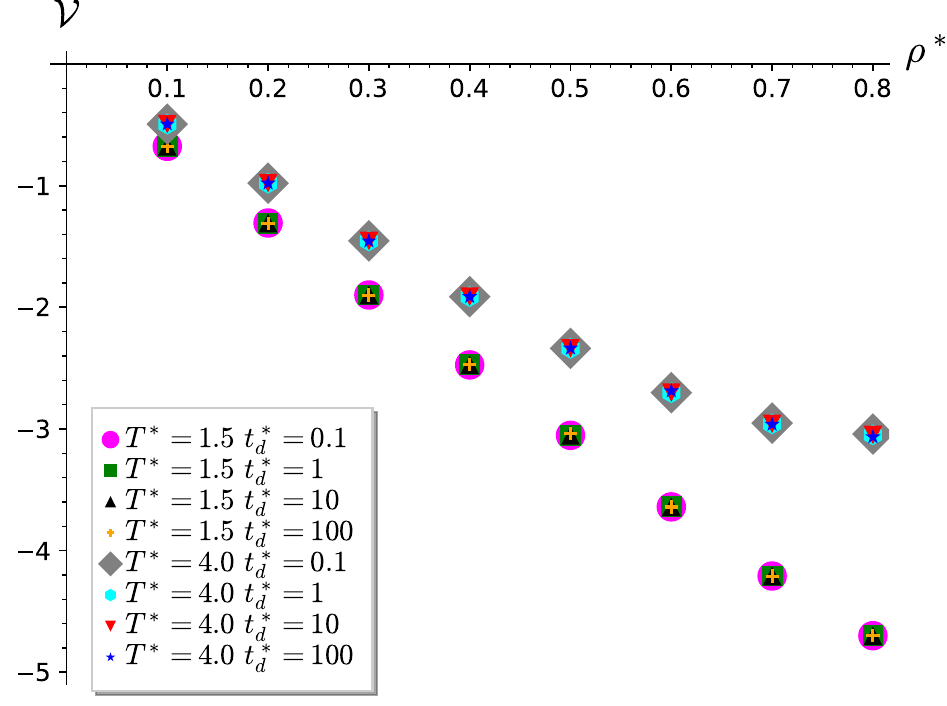}
	\end{center}
	\caption{Dimensionless potential energy per particle $\mathcal{V}$ as a function of the reduced density $\rho^*$ (see Table \ref{tab:1} for a description of reduced quantities) for different values of reduced damping time $t_d^*$, of a system of $N$ particles interacting via pseudo-hard sphere (a) and Lennard Jones (LJ) potential (b). For the LJ potential, results for two values of the reduced temperature, $T^*=1.5$ and $T^*=4$, are shown. A number of particles $N=32000$ was used in the MD simulations. The statistical error of the data is not greater than $0.15\%$ }\label{f.PE}
\end{figure}

The introduction of the thermostat modifies the transport properties; both $\eta$ and $\eta_0$ depend on $t_d$. However, the thermodynamic state remains unchanged. The equation of state is typically represented by the compressibility factor $Z = P/(k_B T \rho)$, where $P$ is the pressure, $k_B$ is Boltzmann constant, $T$ is the temperature and $\rho=N/V$ is the particle density. For given values of $T$ and $\rho$, the compressibility factor is determined by the pressure $P$. Using the virial equation (see, for example, Sec.\ 3.7 in \cite{pathria}), it can be shown that the pressure does not change when the thermostat is introduced (see Sec.\ IV in \cite{marchioni} for more details); therefore, the thermodynamic state is independent of $t_d$. Since $\phi$ is assumed to be a thermodynamic function, it does not depend on $t_d$. To illustrate this point we show in Fig.~\ref{f.PE} numerical results of the potential energy at equilibrium for different values of $t_d$ and different types of pair-wise interactions. We can observe that the noise intensity does not have an effect on the potential energy at equilibrium. 
Consequently, according to Eq.\ \eqref{e.etaphi}, $\eta/\eta_0$ is independent of $t_d$, unlike $\eta$ and $\eta_0$. 

To verify Eq.\ \eqref{e.etaphi}, we present numerical results from molecular dynamics (MD) simulations in Sections \ref{s.hs} and \ref{s.LJ}. These simulations evaluate $\eta/\eta_0$ for both pseudo hard spheres and the Lennard-Jones potential across different values of $t_d$. 
The viscosity in both cases was calculated using the auto-correlation of the stress/pressure tensor integration method. The MD simulations were performed with $108000$ particles for the pseudo-hard sphere potential and $32000$ particles for the Lennard-Jones potential. We used 25 samples with an equilibration run of $10^6$ time steps and a data gathering run of $5\times10^6$, with a time step size of $0.001$.
In the next section, we will first calculate the small concentration viscosity, $\eta_0$, analytically.

\section{Viscosity at small concentration with Langevin thermostat}
\label{s.eta0}

The small concentration viscosity can be analytically obtained using Boltzmann's theory for the under-damped case (large $t_d$) and the Langevin description of Brownian motion for the over-damped case (small $t_d$). As discussed in the following sections, both pseudo-hard sphere and Lennard-Jones (LJ) potentials have a characteristic particle diameter, $\sigma$, and a characteristic energy, $\epsilon$. It is convenient to use reduced (dimensionless) quantities, denoted with an asterisk, which are obtained by scaling with a function of $\sigma$, $\epsilon$ or $m$ (see Table \ref{tab:1}). Note that a large damping time $t_d$ means $t_d^* \gg 1$ and a small $t_d$ corresponds to $t_d^* \ll 1$.

\begin{table}
    \centering
    \begin{tabular}{ccc}
        Quantities & Real units & Reduced quantities\\
        \hline
         Time & $\tau$ & $\tau^{\ast}=\tau\, \sigma^{-1} \sqrt{\epsilon/m}$\\
        Temperature & $T$ & $T^* = k_B T/\epsilon$\\
        Density & $\rho$ & $\rho^*=\rho\sigma^3$ \\
        Viscosity & $\eta$ & $\eta^*=\eta \sigma^{2}/\sqrt{m \epsilon}$ \\
    \end{tabular}
    \caption{Relations between real and reduced (dimensionless) quantities}
    \label{tab:1}
\end{table}

\subsection{Viscosity in the noise-free limit}
For $t_d^* \gg 1$, the influence of the thermostat is negligible. In this case, the viscosity at low concentration can be obtained via the Boltzmann's kinetic theory, with the Chapman-Enskog method \cite{chapman,Hirschfelder},
\begin{equation}\label{e.dboltz}
\eta_B = \frac{5}{16\, \sigma^2\, \Omega_{22}}\sqrt{\frac{m k_B T}{\pi}},
\end{equation}
where $\Omega_{22}$ is the so-called reduced collision integral.
The reduced collision integral for hard spheres (HS) is equal to 1, while for the LJ potential, it has been numerically shown \cite{fokin} that it can be represented by the following function of the reduced temperature:
 \begin{equation}\label{e.Omega22}
 \Omega_{22}^{\rm LJ}(T^*) = \exp\left(-\frac{1}{6} \log(T^*) + \log{\frac{17}{18}} + \sum_{i=0}^{5} a_i {T^*}^{-i/2}\right),
 \end{equation}
with $a_0=0.31081$, $a_1=-0.171211$, $a_2=-0.715805$, $a_3=2.48678$, $a_4=-1.78317$, $a_5=0.394405$. Using the definitions of  dimensionless quantities in table \ref{tab:1}, we have
\begin{equation}\label{e.D0LJ}
 \eta_B^* = \frac{5}{16\, \Omega_{22}}\sqrt{\frac{T^*}{\pi}}.
\end{equation}

In the context of the excess entropy scaling \cite{rosenfeld,rosenfeld2,dyre}, the dimensionless viscosity is obtained in a different way; it is given by $\tilde{\eta} = \eta \, \rho^{-2/3} (m k_B T)^{-1/2}$. The proposal is similar to ours in that a state function, the excess entropy in this case, is also used to represent a transport coefficient. But there are important differences. We propose that the ratio $\eta/\eta_0$ is a function of the thermodynamic state. Excess entropy scaling postulates that $\tilde{\eta}$ is an exponential of the excess entropy with two adjustable parameters, where the so-called macroscopically reduced units are used; this reduced units vary with the thermodynamic state; space, time or energy are scaled with combinations of density, temperature and particle mass. We do not apply any such scaling and do not propose a specific form for $\eta/\eta_0$.

\subsection{Viscosity in a Brownian system}

If $t_d^* \ll 1$ the interaction with the solvent particles (the reservoir) becomes significantly more dominant than the interaction between the larger particles in the system. In this case, the dynamics of the system closely resembles that of a Brownian particle. 
We use the Langevin theory and the Green Kubo formula to derive an expression for the viscosity in this limit.

We consider a low-density system of Brownian particles, where interactions between particles are negligible. We aim to calculate the viscosity of this system. It should not be confused with the viscosity of the medium in which the particles move, which is related to the friction coefficient $\gamma=1/t_d$.

We consider a system with volume $V$, at temperature $T$, in which there are $N$ particles with mass $m$,  positions $\textbf{r}_i=(x_i,y_i,z_i)$ and velocities $\textbf{v}_i=(v_{xi},v_{yi},v_{zi})$, with $i=1\cdots N$. The Green-Kubo formula for the viscosity in an isotropic medium at low concentration is (see, for example, Sec.\ 21.9 in \cite{mcquarrie})
\begin{equation}\label{e.A.5}
   \eta_L = \frac{m^2}{Vk_BT}\sum_{i,j}\int_0^\infty dt\, \langle v_{xi}(0)v_{yi}(0)v_{xj}(t)v_{yj}(t) \rangle.
\end{equation}
Terms containing inter-particle forces are absent since the dilute gas limit is considered. Terms containing forces with the reservoir (friction and noise) are also neglected since it can be shown that this exchange does not contribute to the momentum flux in the system of particles.

For Brownian particles, we consider that components $x$ and $y$ of the velocity are uncorrelated, and there is also no correlation when $i\neq j$. Then, 
\begin{align}\label{e.A.6}
\eta_{L}&=\frac{m^2}{Vk_BT}\sum_i\int_0^\infty dt\, \langle v_{xi}(0)v_{xi}(t) \rangle \langle v_{yi}(0)v_{yi}(t) \rangle\nonumber\\
&=\frac{m^2\rho}{k_BT}\int_0^\infty dt\, \langle v_x(0)v_x(t) \rangle ^2,
\end{align}
where $\rho=N/V$ is the number density, and the isotropy property has been used. 

The auto-correlation function of the velocity of Brownian particles is given by (see, for example, Sec.\ 5.E in \cite{reichl}),
\begin{equation}\label{e.A.7}
 \langle v_x(0)v_x(t) \rangle = \frac{k_BT}{m}e^{-t/t_d}.   
\end{equation}
Using this expression in Eq.~\ref{e.A.6} we obtain
\begin{equation}\label{e.A.8}
\eta_{L}=\frac{\rho\, k_B T\, t_d}{2}.    
\end{equation}
or, using reduced quantities,
\begin{equation}\label{e.dlanga}
\eta_L^*= \frac{\rho^*\, T^*\, t_d^*}{2}.
\end{equation}
This result represents the viscosity of the system of particles, not to be confused, as mentioned before, with the viscosity of the solvent, that is related to the damping time. In fact, the damping time is inversely proportional to the viscosity of the solvent; therefore, according to Eq.\ \eqref{e.A.8}, $\eta_L$ is also inversely proportional to the viscosity of the solvent. 

\subsection{General form of the viscosity at small concentration}

For intermediate values of $t_d^*$, we employ a similar approach to that presented in Ref.\ \cite{marchioni} for the self-diffusion coefficient. 
Since viscosity measures the capacity for momentum transfer between different parts of a fluid, its inverse represents resistance to this transfer.
Momentum transfer can be described using Newton's law of viscosity. For example, in laminar flow in the $x$ direction, $\tau_{xy}=\eta\, \frac{\partial{u}}{\partial{y}}$, where $\tau_{xy}$ is the shear stress (the force per unit area in the $x$ direction, also interpreted as the momentum density current in the $y$ direction), and $u$ is the average velocity of the fluid in the $x$ direction. This can be compared to Ohm's law in an electrical circuit, that is, $I = R^{-1}\mathcal{V}$, where $I$ is the electric current, $\mathcal{V}$ is the voltage and $R$ is the electric resistance. In this comparison, we make the following correspondences: $I\rightarrow \tau$,  $\mathcal{V}\rightarrow \frac{\partial{u}}{\partial{y}}$, and $R\rightarrow \eta^{-1}$. By assuming that momentum flux is analogous to the electric current in a series circuit and considering the equivalent resistance, we can derive an expression for the low concentration viscosity:
\begin{equation}\label{e.D0}
\frac{1}{\eta_0^*} = \frac{1}{\eta_B^*} + \frac{1}{\eta_L^*},
\end{equation}
where $1/\eta_B^*$ and $1/\eta_L^*$ are the resistances due to collisions with other particles and due to the Langevin thermostat, respectively. Using Eqs.\ \eqref{e.D0LJ} and \eqref{e.dlanga} we have,
\begin{equation}\label{e.D02}
\eta_0^* = \frac{T^*\, t_d^*\, \rho^*}{16/5\, \rho^*\, t_d^* \, \Omega_{22} \sqrt{T^*\, \pi} + 2}.
\end{equation}
It is straightforward to verify that $\eta_0^* \simeq \eta_L^*$ for $t_d\ll 1$ and that $\eta_0^* \simeq \eta_B^*$ for $t_d \gg 1$.

Let us notice that $\eta_0$ is a function of the concentration $\rho$. The ratio $\eta/\eta_0$ that we calculate below, should be understood as the viscosity divided by the expression for viscosity which holds at small concentration, and which is a function of concentration; this function can be evaluated at values of $\rho$ where it does not hold by considering an ideal hypothetical system, where the condition of small interactions always holds. Similarly, we can calculate the ideal pressure $P_0$ using the ideal gas equation for densities where it does not hold, and write the equation of state as usual, as $P/P_0 = Z$, where $Z$ is the compressibility factor. Another example is the Enskog diffusion coefficient, $D_E$, for hard spheres (without noise); it is written as $D_E/D_0 = 1/g$, where $g$ is the radial distribution function at contact, and $D_0$ behaves as $1/\rho$.

\begin{figure}
	\begin{center}
		\includegraphics[width=\linewidth]{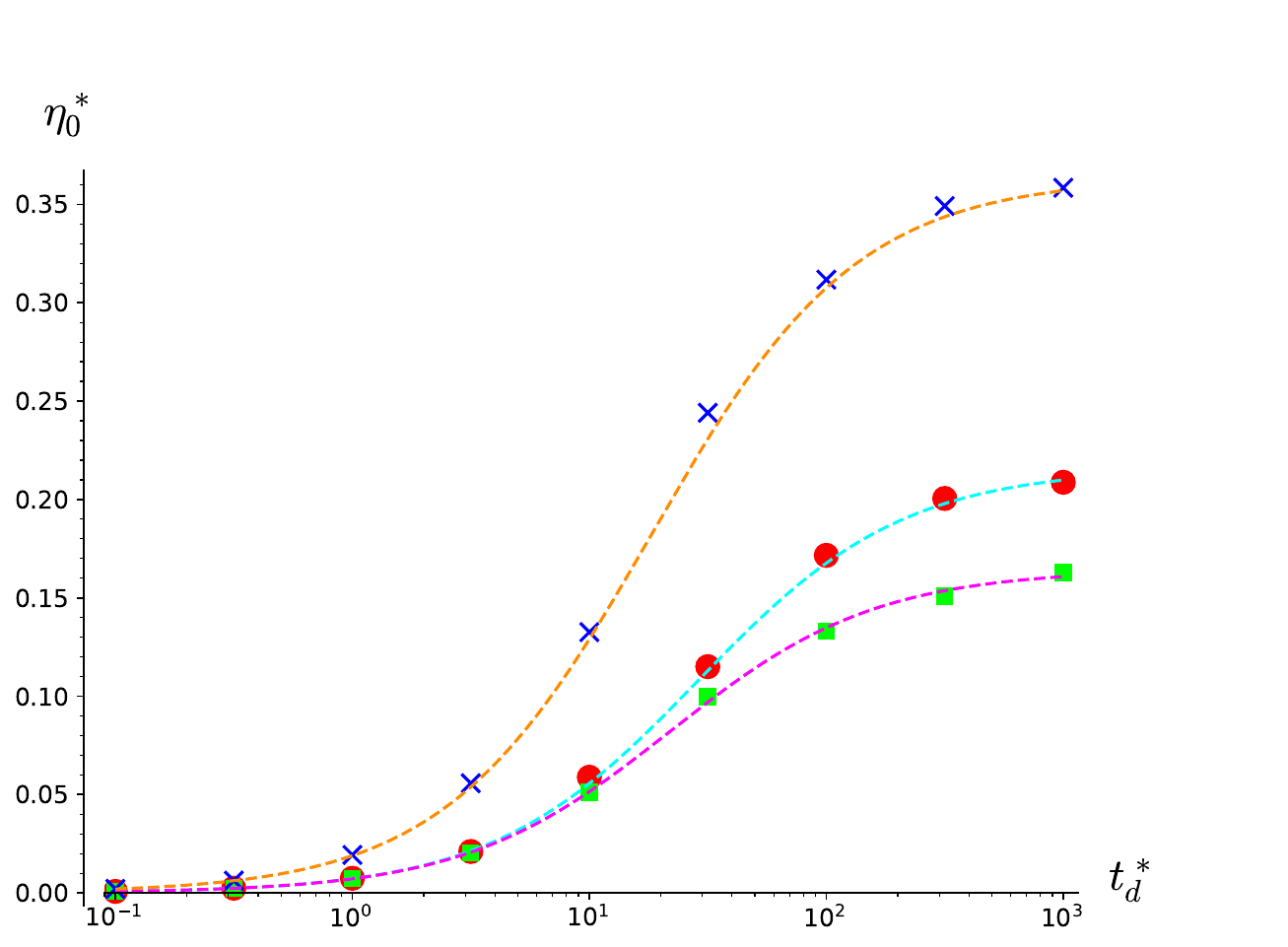}
	\end{center}
	\caption{Reduced viscosity at small concentration, $\eta_0^*$, against damping time, $t_d^*$, for reduced density $\rho^*=0.01$. The results for pseudo-hard spheres (red circles), Lennard Jones with temperature $T^\ast=1.5$ (green boxes) and $T^\ast=4$ (blue crosses) are shown. The dashed lines represent Eq.\ \eqref{e.D02}. The maximum statistical error of the data is $3\%$.
 }\label{f.D0}
\end{figure}

In Fig.~\ref{f.D0} we compare Eq.~\eqref{e.D02} with numerical simulations performed at low concentration for pseudo-hard spheres ($\Omega^{\text{HS}}_{22}=1$) and for the LJ potential, where $\Omega^{\text{LJ}}_{22}$ is given by Eq.\ \eqref{e.Omega22}. We observe that there is a good agreement between theory and simulations.

\section{Pseudo Hard Spheres}
\label{s.hs}

The pairwise potential for hard-spheres (HS) is
\begin{equation}
u_{\text{HS}}=\left\{ \begin{array}{cc}  
\infty & \ \ \ r < \sigma  \\[\medskipamount]
0 & \ \ \ r \ge \sigma 
\end{array}  \right. ,
\end{equation}
where $\sigma$ is the diameter of the spheres and $r$ is the distance between the centers of the spheres. It is well known that such interactions cannot be successfully implemented in Molecular Dynamics simulations, as particles that come very close to each other would experience an infinite force. However, it is possible to approximately model a hard-sphere fluid using Molecular Dynamics. It has been demonstrated that the Mie potential can effectively reproduce the properties of a hard sphere fluid \cite{jover,Pousaneh}. The potential is given by
\begin{equation}\label{e.pseudo}
u(r) = \left\{ \begin{array}{cc}
\frac{\lambda_r}{\lambda_r - \lambda_a} (\frac{\lambda_r}{\lambda_a})^{\frac{\lambda_a}{\lambda_r - \lambda_a}} \epsilon \left[(\frac{\sigma}{r})^{\lambda_r} - (\frac{\sigma}{r})^{\lambda_a}\right] + \epsilon & \ \ \ r < \sigma (\frac{\lambda_r}{\lambda_a})^{\frac{1}{\lambda_r - \lambda_a}} \\[\medskipamount]
0 & \ \ \ r \ge \sigma (\frac{\lambda_r}{\lambda_a})^{\frac{1}{\lambda_r - \lambda_a}}
\end{array}
  \right. ,
\end{equation}
with $\lambda_r=50$ and $\lambda_a=49$. Note that this potential generalizes the WCA potential \cite{weeks}, which itself is a cut and shifted version of the well-known Lennard Jones potential. In this context, $\lambda_r$ and $\lambda_a$ are the exponents for the repulsive and attractive terms, respectively, while $\epsilon$ represents the depth of the attractive well. It has been shown that to accurately reproduce the properties of a hard-sphere fluid, the reduced temperature must be set at $T^*=1.5$. Since this potential does not exactly replicate a hard-core interaction, it is typically referred to as a pseudo hard-sphere potential \cite{jover}.

\begin{figure}
	\begin{center}
		\includegraphics[width=\linewidth]{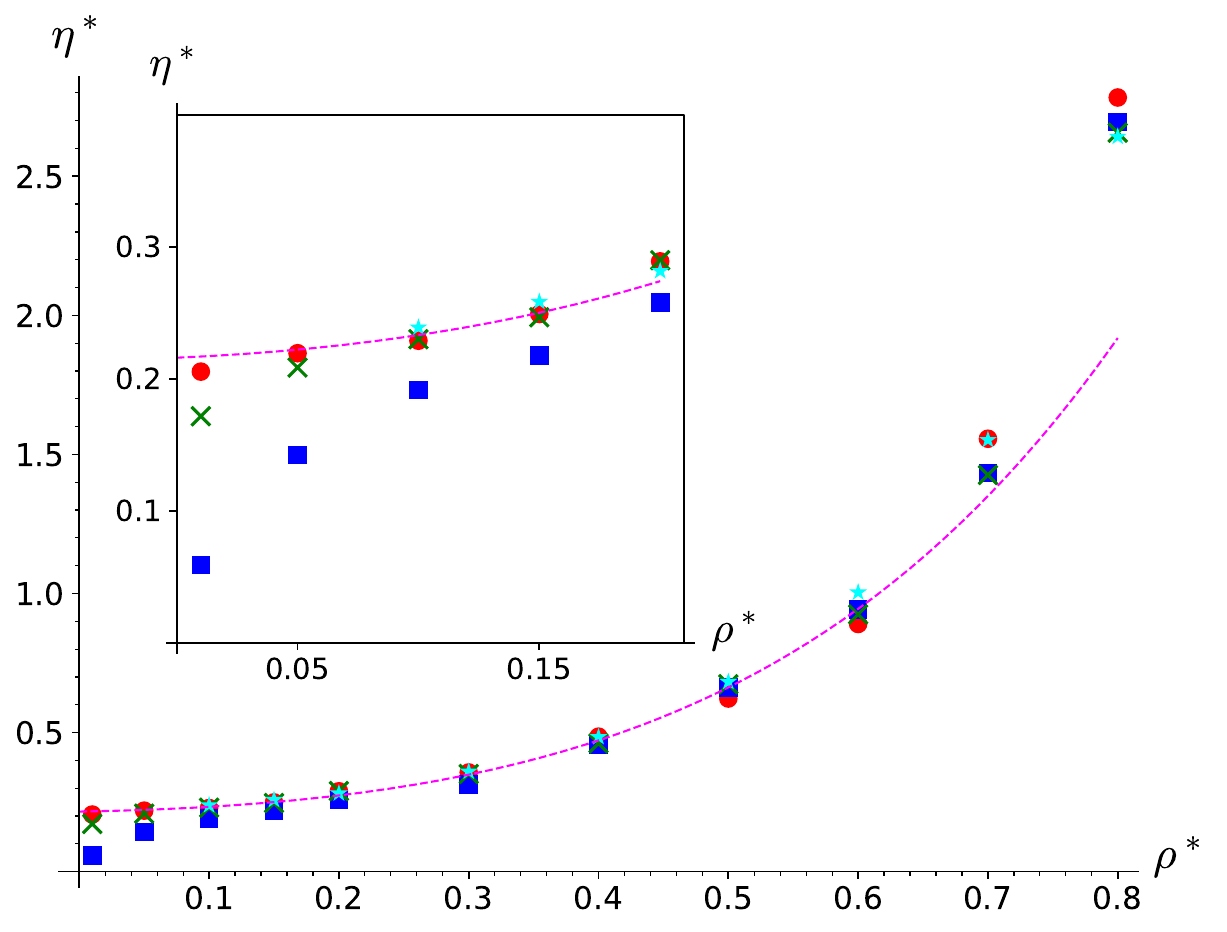}
	\end{center}
 	\caption{Viscosity for pseudo hard-spheres ($T^*=1.5$), $\eta^\ast$, versus reduced density, $\rho^*$, for different values of damping time $t_d^*$. The red circles are noise-free results (with the noise-free case corresponding to the limit of infinite $t_d$), blue boxes correspond to $t_d^*=10$ and green crosses to $t_d^*=100$. The maximum statistical error of the data is $5\%$. The cyan stars represent numerical results from Pieprzyk \textit{et al.} \cite{pieprzyk2} and the dashed line corresponds to the theoretical proposal of Enskog \cite{enskog}. The inset shows the low-desity region.}
 \label{f.D_eta}
\end{figure}

We have examined how viscosity varies with damping time at low concentration; see Eq.~\eqref{e.D02} and Fig.~\ref{f.D0}. The results indicate that, as the coupling with the reservoir decreases (i.e.\ as $t_d$ increases), the viscosity rises. At higher density, Fig.~\ref{f.D_eta} illustrates the reduced viscosity as a function of reduced density for three different values of damping time (with the noise-free case corresponding to the limit of infinite $t_d$). As anticipated, for a given density, viscosity increases with larger $t_d$. The noise-free case, represented by red circles (our results) or cyan stars (from Ref.\ \cite{pieprzyk2}) in Fig.\ \ref{f.D_eta}, is well described by the Enskog theory \cite{enskog} for small and moderate densities; see also Eq.\ (9.25) in Ref.\ \cite{mulero1}.


Our main hypothesis is that $\eta/\eta_0=\phi$ is a state function. Therefore, $\phi$ should depend solely on the thermodynamic state of the system, which, as explained in Sec.\ \ref{s.hyp}, is unaffected by the presence of noise. At low densities $\phi \simeq 1$, and it increases with concentration. However, for fixed density, changing the damping time should not modify $\phi$.
\begin{figure}
	\begin{center}
		\includegraphics[width=\linewidth]{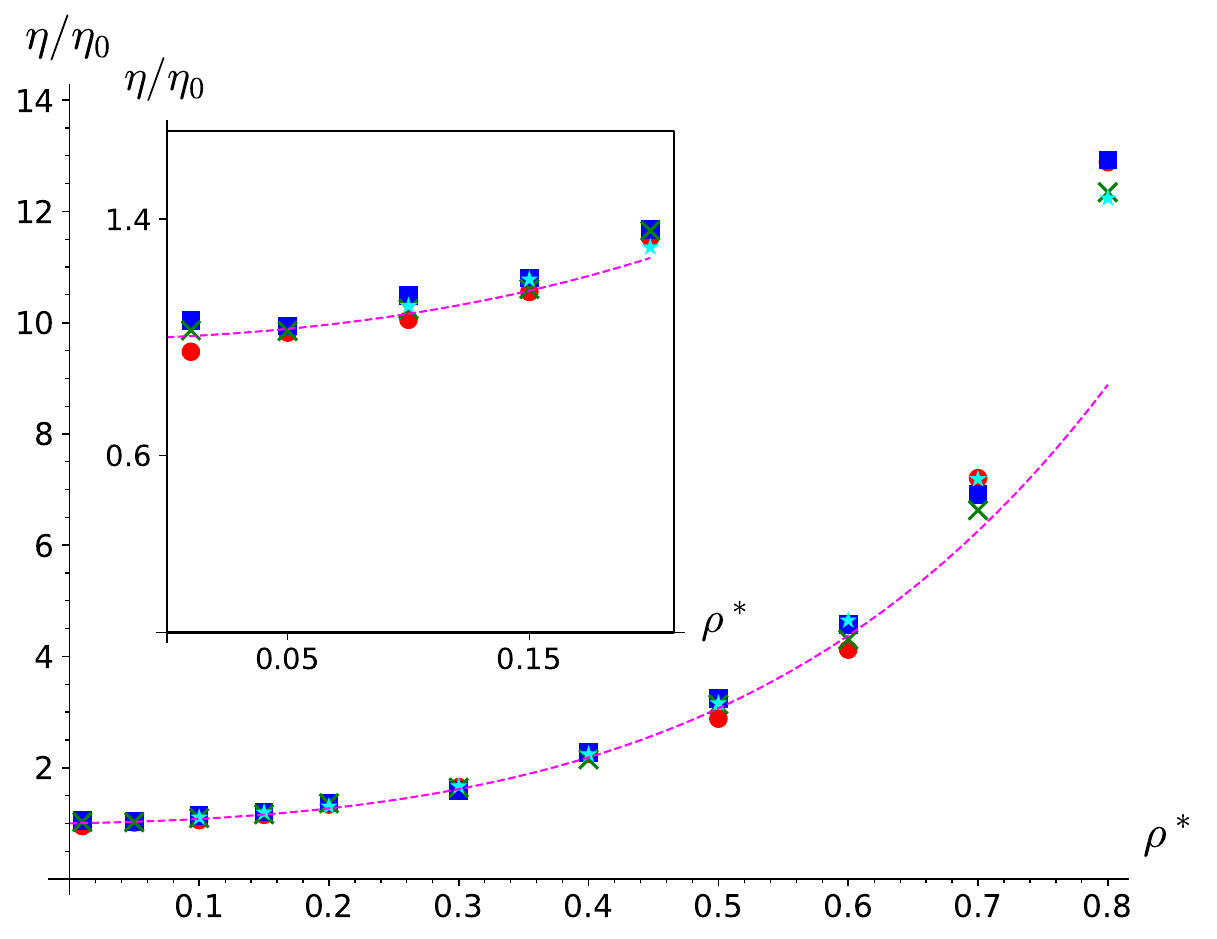}
	\end{center}
	\caption{Viscosity divided by low-concentration viscosity for pseudo hard-spheres ($T^*=1.5$), $\eta/\eta_0$, versus reduced density, $\rho^*$. The red circles are noise-free results (with the noise-free case corresponding to the limit of infinite $t_d$), blue boxes correspond to $t_d^*=10$ and green crosses to $t_d^*=100$. The maximum statistical error of the data is $5\%$. The cyan stars are the numerical results from Pieprzyk \textit{et al.} \cite{pieprzyk2} and the dashed line corresponds to the theoretical proposal of Enskog \cite{enskog}. The inset zooms in on the low-desity region where, when compared with Fig.\ \ref{f.D_eta}, a collapse of the curves is observed.}\label{f.D_etaeta0}
\end{figure}

In Fig.~\ref{f.D_etaeta0} we present simulation results of $\eta/\eta_0 = \phi$ as a function of concentration for various damping times. We find that $t_d$ does not influence the behavior of $\eta/\eta_0$. The values of $\eta/\eta_0$ for different damping times $t_d^*$ agree within numerical errors. For comparison, we also include numerical results from Pieprzyk \textit{et al.} \cite{pieprzyk2} for hard spheres without noise ($t_d \rightarrow \infty$)  and the theoretical proposal of Enskog \cite{enskog}, that can also be found as Eq.\ (9.25) of Ref.\ \cite{mulero1}.

\section{Lennard-Jones potential}
\label{s.LJ}

In this section we consider a pairwise interaction between particles modeled by the Lennnard-Jones potential
\begin{equation}
	u_{LJ}(r) = 4\epsilon \left[ \left( \frac{\sigma}{r} \right)^{12} - \left( \frac{\sigma}{r} \right)^6 \right],
\end{equation}
where $\epsilon$ represents the depth of the potential well, $\sigma$ is the distance at which the potential is zero and, for the simulations, the
potential was cut and shifted at radius 2.5$\sigma$. Our goal is to investigate whether $\eta/\eta_0$ also behaves as a state function for a Lennard Jones fluid.

\begin{figure}
	\begin{center}
	a)\includegraphics[width=7.5cm]{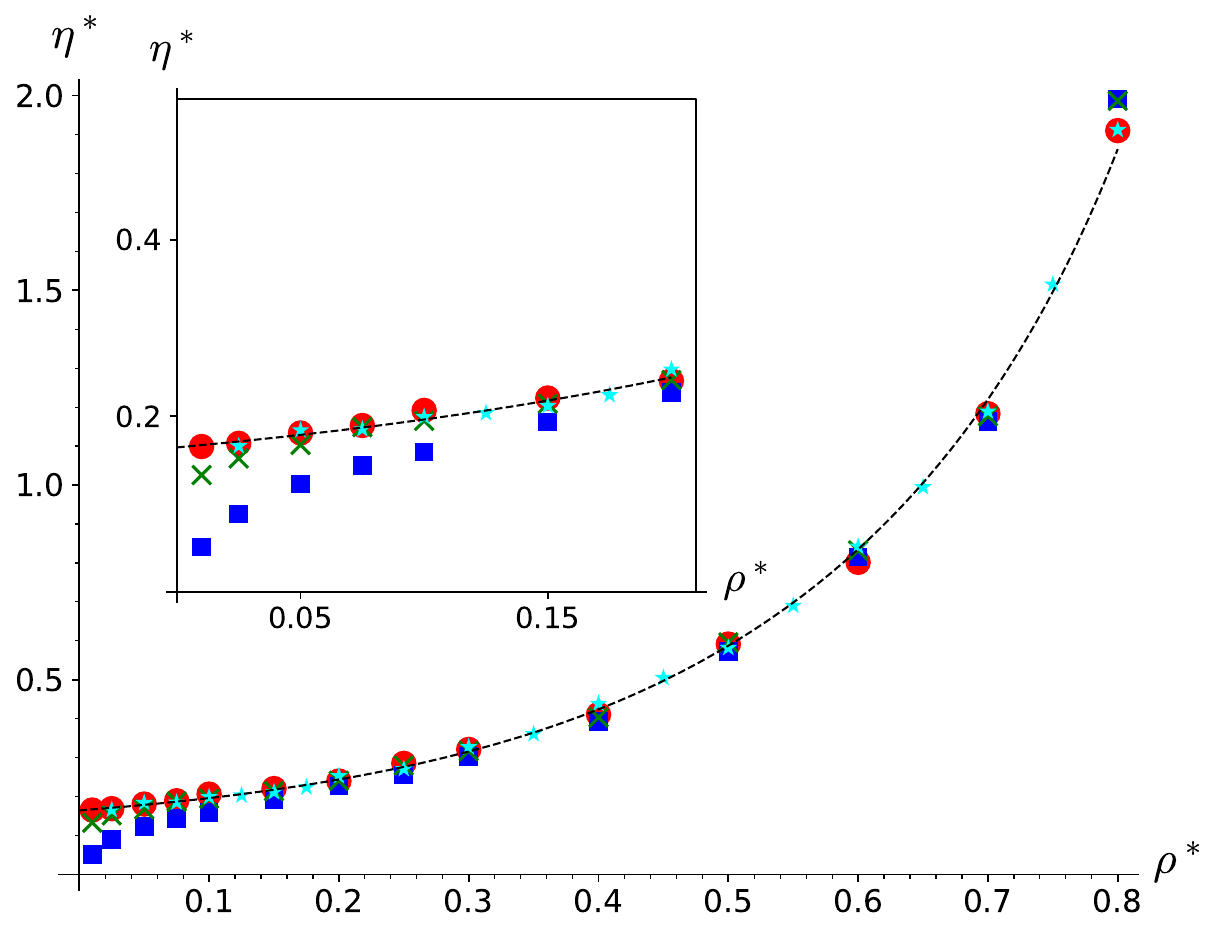}
        b)\includegraphics[width=7.5cm]{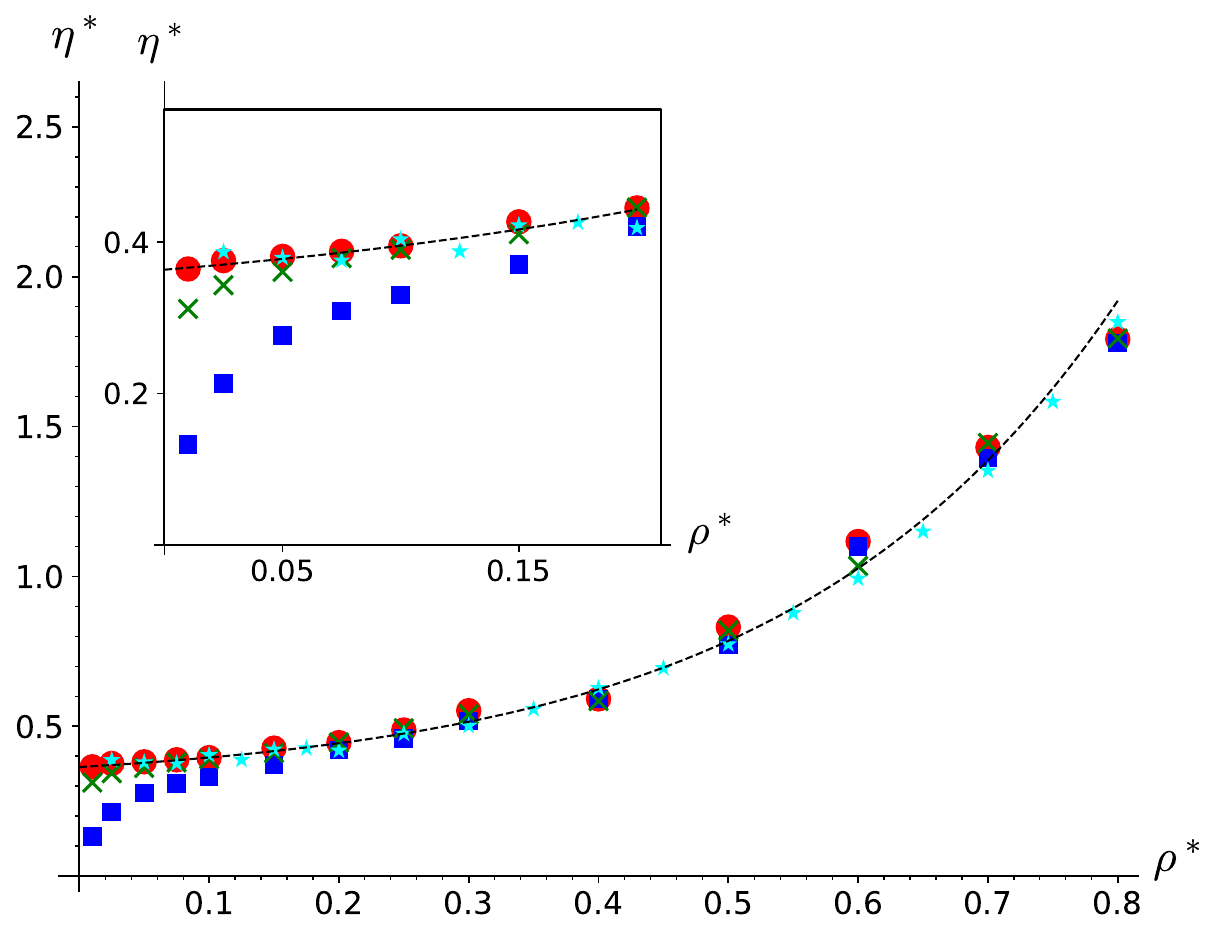}
	\end{center}
	\caption{Reduced viscosity $\eta^\ast$ for the Lennard-Jones potential as a function of the reduced density $\rho^\ast$, for temperatures (a) $T^*=1.5$ and (b) $T^*=4$. The red circles are noise-free results ($t_d \to \infty$), blue boxes correspond to $t_d^*=10$ and green crosses to $t_d^*=100$. The maximum statistical error of the data is $5\%$. The cyan stars represent the numerical results (without noise) from Meier \cite{meier3} and the dashed line corresponds to the theoretical proposal of Galli\'ero et al. \cite{Galliero}. The inset illustrates the low-desity region.}\label{f.D08}
\end{figure}

To test this, we first present in Fig.~\ref{f.D08} the results of the simulations showing the reduced viscosity as a function of the concentration for three different noise intensities and two different temperatures. The results indicate that viscosity is proportional to both density and damping time.

\begin{figure}
	\begin{center}
	a)\includegraphics[width=7.5cm]{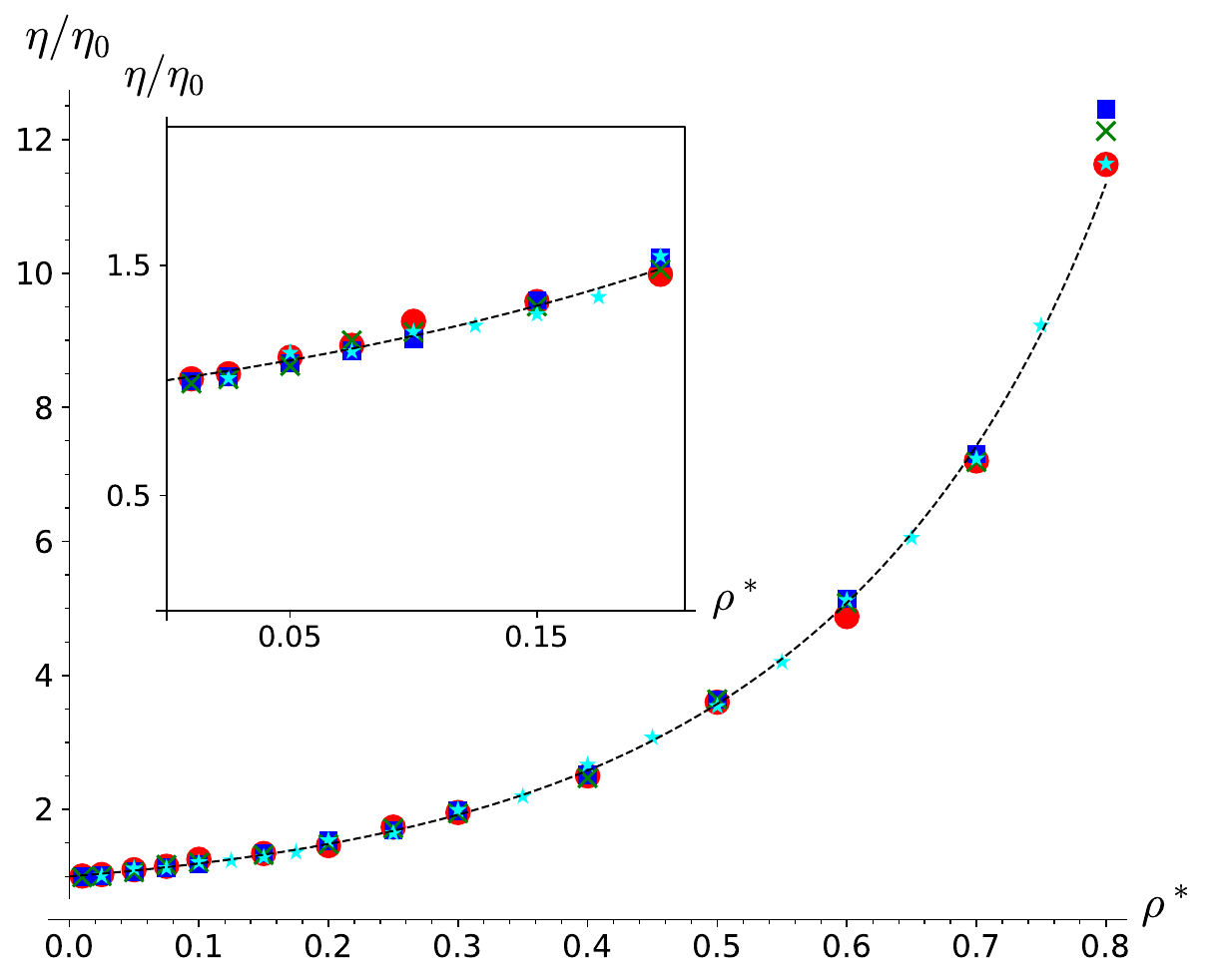}
        b)\includegraphics[width=7.5cm]{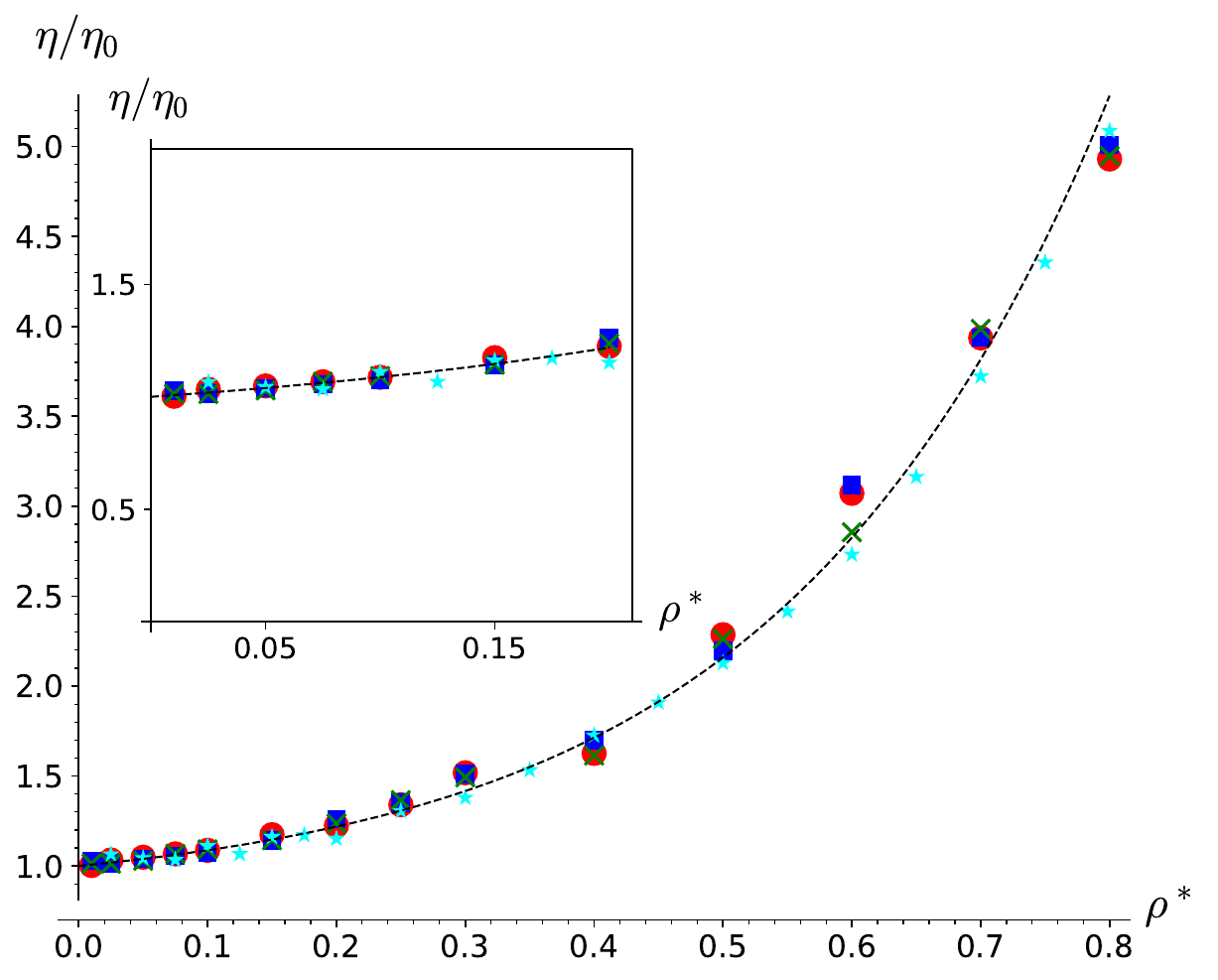}
	\end{center}
	\caption{Viscosity divided by viscosity at small concentration, $\eta/\eta_0$, as a function of the reduced density $\rho^\ast$ for the Lennard-Jones potential, with temperatures (a) $T^*=1.5$ and (b) $T^*=4$. The red circles are noise-free results ($t_d \to \infty$), blue boxes correspond to $t_d^*=10$ and green crosses to $t_d^*=100$. The maximum statistical error of the data is $5\%$. The cyan stars represent the noise-free numerical results from Meier \cite{meier3} and the dashed line corresponds to the theoretical proposal of Galli\'ero et al. \cite{Galliero}. The inset highlights the low-desity region; comparing with Fig.\ \ref{f.D08}, the collapse of the curves is observed.}\label{f.D09}
\end{figure}

Next, in Fig.~\ref{f.D09} we divide the viscosity, $\eta$, by the viscosity at low concentration, $\eta_0$, as given by Eq.~\eqref{e.D02}; that is, $\eta/\eta_0=\phi$. The curves for different damping times collapse, a result that is consistent with the hypothesis that $\phi$ depends solely on the thermodynamic state of the system. 
 
\section{Conclusions}
\label{s.conclusions}

In this work, we investigated the hypothesis that the shear viscosity of dense fluids can be expressed as $\eta = \phi\, \eta_0$, where $\eta_0$ is the low-density viscosity and $\phi$ is a state function. 

We developed a method to test this hypothesis by coupling the fluid to a Langevin thermostat characterized by a damping time $t_d$. This allowed us to modify $\eta$ and $\eta_0$ without altering the thermodynamic state of the system.
An analytical expression for $\eta_0$ as a function of $t_d$ was derived by combining Boltzmann and Langevin theories. This expression showed good agreement with numerical simulations for both pseudo hard spheres and Lennard-Jones fluids.
Molecular dynamics simulations were performed for pseudo hard spheres and Lennard-Jones fluids over a range of densities and damping times. The results demonstrated that while $\eta$ depends on $t_d$, the ratio $\eta/\eta_0$ is independent of $t_d$ within numerical uncertainty.
This invariance of $\eta/\eta_0$ with respect to $t_d$ supports our hypothesis that $\phi$ is indeed a state function, as it depends only on the thermodynamic state of the system and not on microscopic details. Similar results were obtained for the ratio of the self-diffusion coefficient, $D/D_0$, in Ref.\ \cite{marchioni}. 
The behavior was consistent for both pseudo hard spheres and Lennard-Jones potentials at different temperatures, suggesting the relation may be general for simple fluids.

These results provide evidence for the proposed factorization of viscosity into a low-density term and a thermodynamic function. This formulation could simplify the development of general transport theories for dense fluids by separating the effects of dilute gas kinetics from the thermodynamic consequences of particle interactions.

The hypothesis that $\eta/\eta_0$ (or $D/D_0$) is a state function is consistent with and strengthened by the present numerical results, but it is not proven. Further research, both numerical and analytical, is needed to fully validate or refute the proposition that, in principle, applies to any transport coefficient. This hypothesis, if true, has implications for understanding transport phenomena, as it separates the effects of dilute gas kinetics from the thermodynamic consequences of particle interactions. Traditional methods for calculating transport coefficients often focus on microscopic details, such as particle size, mass, mean free path, or the radial distribution function. Our hypothesis, on the other hand, posits that information at the thermodynamic level is sufficient to obtain the ratio of transport coefficients; it could simplify the development of transport theories for dense fluids by focusing on macroscopic thermodynamic properties rather than microscopic details.

It is important to point out that the analysis conducted across this manuscript is limited to laminar flows, for which there is a clear relation between the shear viscosity and the momentum flux. Even though there are expressions in the literature for calculating the eddy viscosity, inherent to turbulent flows (see Ch.~4 in \cite{Bird}), its study would require the inclusion of hydrodynamic forces, which would significantly increase the complexity of the problem. In this manuscript we showed that $\phi$ is a thermodynamic function for a simple interacting system of particles. The next natural step would be to verify if this statement holds for other pairwise interactions, or for more complex systems.

\section*{Acknowledgments}
This work was partially supported by Consejo Nacional de Investigaciones
Cient\'ificas y T\'ecnicas (CONICET, Argentina, PUE 22920200100016CO).

\appendix
\section{Details on the Molecular Dynamics simulations}\label{Ap.1}

The Molecular Dynamics simulations were conducted using LAMMPS code. Initially, a simulation box with periodic boundary conditions is created, and then the atoms are placed in an fcc lattice, each atom has a random initial velocity with a Maxwellian distribution at temperature $T^*$. Next, a thermalization stage begins, but first a Langevin thermostat is introduced (the corresponding command in LAMMPS script is ``fix langevin"). This thermostat adds a viscous force $F_f$ proportional to the particle's velocity $v$ and a random force $F_r$, with
\[
F_f(t)=-\frac{m}{t_d}v(t)
\]
and 
\[
F_r(t)=\sqrt{\frac{\kappa_B T m}{ t_d\;\Delta t}}\xi(t),
\]
where $\Delta t$ is the time step of the simulation and $\xi(t)$ is a random variable normally distributed and delta correlated. This proportionality relation is derived from the fluctuation-dissipation theorem.

The numerical integration of the equations of motion is performed using the velocity-Verlet algorithm (``fix nve" command). This algorithm updates the position $\textbf{r}$ and velocity $\textbf{v}$ of the particles as follows:
\begin{align}
\textbf{r}(t+\Delta t)&=\textbf{r}(t)+\textbf{v}(t)\Delta t+\frac{\Delta t^2}{2m}\textbf{f}(t)\nonumber,\\
\textbf{v}(t+\Delta t)&=\textbf{v}(t)+\frac{\Delta t}{2m}[\textbf{f}(t)+\textbf{f}(t+\Delta t)],
\end{align}
where $\textbf{f}(t)$ is the total force that acts upon a particle at time $t$. 


We clarify that in the simulations Lennard Jones units were used, for which all quantities are scaled with combinations of the fundamental parameters $\sigma$, $\epsilon$ and $m$, as illustrated in Table \ref{tab:1}. For instance, the position has units of $\sigma$, the velocity units of $\sqrt{\epsilon/m}$, and the force units of $\epsilon/\sigma$.

We used an integration timestep $\Delta t^\ast=0.001$, $t^\ast=t\sqrt{m\sigma^2/\epsilon}$. The minimum value of the damping time $t_d$ that we used is $0.1$, that is, two orders of magnitude larger than the integration timestep; this assures a good enough time resolution for the calculation of time correlations needed to evaluate transport coefficients.

The viscosity is calculated via the Green-Kubo formula. After the thermalization stage ($10^6$ time steps), the time-autocorrelation of the off-diagonal elements of the stress tensor are computed using the following parameters: sample values are taken every $N_e=10$ time steps, $N_r=4000$ samples are used to compute the correlation, and the correlation is calculated every $N_f=N_e N_r=40000$ timesteps (``fix ave/correlate" command). The data gathering run lasts $5\times 10^6$ time steps.
The global potential energy, illustrated in Fig.~\ref{f.PE}, was calculated taking the sum over particle pair interactions (``compute pe" command).

\bibliography{Visco.bib}

\begin{thebibliography}{40}%
\makeatletter
\providecommand \@ifxundefined [1]{%
 \@ifx{#1\undefined}
}%
\providecommand \@ifnum [1]{%
 \ifnum #1\expandafter \@firstoftwo
 \else \expandafter \@secondoftwo
 \fi
}%
\providecommand \@ifx [1]{%
 \ifx #1\expandafter \@firstoftwo
 \else \expandafter \@secondoftwo
 \fi
}%
\providecommand \natexlab [1]{#1}%
\providecommand \enquote  [1]{``#1''}%
\providecommand \bibnamefont  [1]{#1}%
\providecommand \bibfnamefont [1]{#1}%
\providecommand \citenamefont [1]{#1}%
\providecommand \href@noop [0]{\@secondoftwo}%
\providecommand \href [0]{\begingroup \@sanitize@url \@href}%
\providecommand \@href[1]{\@@startlink{#1}\@@href}%
\providecommand \@@href[1]{\endgroup#1\@@endlink}%
\providecommand \@sanitize@url [0]{\catcode `\\12\catcode `\$12\catcode
  `\&12\catcode `\#12\catcode `\^12\catcode `\_12\catcode `\%12\relax}%
\providecommand \@@startlink[1]{}%
\providecommand \@@endlink[0]{}%
\providecommand \url  [0]{\begingroup\@sanitize@url \@url }%
\providecommand \@url [1]{\endgroup\@href {#1}{\urlprefix }}%
\providecommand \urlprefix  [0]{URL }%
\providecommand \Eprint [0]{\href }%
\providecommand \doibase [0]{http://dx.doi.org/}%
\providecommand \selectlanguage [0]{\@gobble}%
\providecommand \bibinfo  [0]{\@secondoftwo}%
\providecommand \bibfield  [0]{\@secondoftwo}%
\providecommand \translation [1]{[#1]}%
\providecommand \BibitemOpen [0]{}%
\providecommand \bibitemStop [0]{}%
\providecommand \bibitemNoStop [0]{.\EOS\space}%
\providecommand \EOS [0]{\spacefactor3000\relax}%
\providecommand \BibitemShut  [1]{\csname bibitem#1\endcsname}%
\let\auto@bib@innerbib\@empty
\bibitem [{\citenamefont {Chapman}\ and\ \citenamefont
  {Cowling}(1970)}]{chapman}%
  \BibitemOpen
  \bibfield  {author} {\bibinfo {author} {\bibfnamefont {S.}~\bibnamefont
  {Chapman}}\ and\ \bibinfo {author} {\bibfnamefont {T.~G.}\ \bibnamefont
  {Cowling}},\ }\href@noop {} {\emph {\bibinfo {title} {The Mathematical Theory
  of Non-Uniform Gases}}},\ \bibinfo {edition} {3rd}\ ed.\ (\bibinfo
  {publisher} {Cambridge University Press},\ \bibinfo {year}
  {1970})\BibitemShut {NoStop}%
\bibitem [{\citenamefont {Enskog}(1922)}]{enskog}%
  \BibitemOpen
  \bibfield  {author} {\bibinfo {author} {\bibfnamefont {D.}~\bibnamefont
  {Enskog}},\ }\bibfield  {title} {\enquote {\bibinfo {title} {Kinetische
  theorie der wärmeleitung, reibung und selbstdiffusion in gewissen
  verdichteten gasen und flüssigkeiten},}\ }\href@noop {} {\bibfield
  {journal} {\bibinfo  {journal} {Kungl. Svenska Vetensk. Handl.}\ }\textbf
  {\bibinfo {volume} {63}},\ \bibinfo {pages} {1} (\bibinfo {year}
  {1922})}\BibitemShut {NoStop}%
\bibitem [{\citenamefont {Koo}\ and\ \citenamefont {Hess}(1987)}]{Koo}%
  \BibitemOpen
  \bibfield  {author} {\bibinfo {author} {\bibfnamefont {Hyearn-Maw}\
  \bibnamefont {Koo}}\ and\ \bibinfo {author} {\bibfnamefont {Siegfried}\
  \bibnamefont {Hess}},\ }\bibfield  {title} {\enquote {\bibinfo {title} {The
  divergence of the viscosity of a fluid of hard spheres as an indicator for
  the fluid-solid phase transition},}\ }\href@noop {} {\bibfield  {journal}
  {\bibinfo  {journal} {Z. Naturforsch}\ }\textbf {\bibinfo {volume} {42a}},\
  \bibinfo {pages} {231--235} (\bibinfo {year} {1987})}\BibitemShut {NoStop}%
\bibitem [{\citenamefont {Hanley}\ \emph {et~al.}(1972)\citenamefont {Hanley},
  \citenamefont {McCarty},\ and\ \citenamefont {Cohen}}]{hanley}%
  \BibitemOpen
  \bibfield  {author} {\bibinfo {author} {\bibfnamefont {H.J.M.}\ \bibnamefont
  {Hanley}}, \bibinfo {author} {\bibfnamefont {R.D.}\ \bibnamefont {McCarty}},
  \ and\ \bibinfo {author} {\bibfnamefont {E.G.D.}\ \bibnamefont {Cohen}},\
  }\bibfield  {title} {\enquote {\bibinfo {title} {Analysis of the transport
  coefficients for simple dense fluid: Application of the modified enskog
  theory},}\ }\href@noop {} {\bibfield  {journal} {\bibinfo  {journal}
  {Physica}\ }\textbf {\bibinfo {volume} {60}},\ \bibinfo {pages} {322--356}
  (\bibinfo {year} {1972})}\BibitemShut {NoStop}%
\bibitem [{\citenamefont {Dymond}(1974{\natexlab{a}})}]{dymond}%
  \BibitemOpen
  \bibfield  {author} {\bibinfo {author} {\bibfnamefont {J.~H.}\ \bibnamefont
  {Dymond}},\ }\bibfield  {title} {\enquote {\bibinfo {title} {Corrected
  {E}nskog theory and the transport coefficients of liquids},}\ }\href@noop {}
  {\bibfield  {journal} {\bibinfo  {journal} {J. Chem. Phys.}\ }\textbf
  {\bibinfo {volume} {60}},\ \bibinfo {pages} {969} (\bibinfo {year}
  {1974}{\natexlab{a}})}\BibitemShut {NoStop}%
\bibitem [{\citenamefont {Dymond}(1974{\natexlab{b}})}]{dymond2}%
  \BibitemOpen
  \bibfield  {author} {\bibinfo {author} {\bibfnamefont {J.~H.}\ \bibnamefont
  {Dymond}},\ }\bibfield  {title} {\enquote {\bibinfo {title} {The
  interpretation of transport coefficients on the basis of the van der waals
  model},}\ }\href@noop {} {\bibfield  {journal} {\bibinfo  {journal}
  {Physica}\ }\textbf {\bibinfo {volume} {75}},\ \bibinfo {pages} {100--114}
  (\bibinfo {year} {1974}{\natexlab{b}})}\BibitemShut {NoStop}%
\bibitem [{\citenamefont {Dymond}(1985)}]{dymond3}%
  \BibitemOpen
  \bibfield  {author} {\bibinfo {author} {\bibfnamefont {J.~H.}\ \bibnamefont
  {Dymond}},\ }\bibfield  {title} {\enquote {\bibinfo {title} {Hard-sphere
  theories of transport properties},}\ }\href@noop {} {\bibfield  {journal}
  {\bibinfo  {journal} {Chem. Soc. Rev.}\ }\textbf {\bibinfo {volume} {14}},\
  \bibinfo {pages} {317--356} (\bibinfo {year} {1985})}\BibitemShut {NoStop}%
\bibitem [{\citenamefont {Dymond}\ and\ \citenamefont
  {Assael}(1996)}]{Dymond1996}%
  \BibitemOpen
  \bibfield  {author} {\bibinfo {author} {\bibfnamefont {J.~H.}\ \bibnamefont
  {Dymond}}\ and\ \bibinfo {author} {\bibfnamefont {M.~J.}\ \bibnamefont
  {Assael}},\ }\bibfield  {title} {\enquote {\bibinfo {title} {Modified
  hard-spheres scheme},}\ }in\ \href@noop {} {\emph {\bibinfo {booktitle}
  {Transport Properties of Fluids: Their Correlation, Prediction and
  Estimation}}},\ \bibinfo {editor} {edited by\ \bibinfo {editor}
  {\bibfnamefont {J.}~\bibnamefont {Millat}}, \bibinfo {editor} {\bibfnamefont
  {J.~H.}\ \bibnamefont {Dymond}}, \ and\ \bibinfo {editor} {\bibfnamefont
  {C.~A.~Nieto}\ \bibnamefont {de~Castro}}}\ (\bibinfo  {publisher} {Cambridge
  University Press},\ \bibinfo {address} {New York},\ \bibinfo {year} {1996})\
  pp.\ \bibinfo {pages} {461--463}\BibitemShut {NoStop}%
\bibitem [{\citenamefont {Silva}\ \emph {et~al.}(2003)\citenamefont {Silva},
  \citenamefont {Coelho}, \citenamefont {Tavares},\ and\ \citenamefont
  {Cardoso}}]{SilvaCoelho}%
  \BibitemOpen
  \bibfield  {author} {\bibinfo {author} {\bibfnamefont {F.~D.~C.}\
  \bibnamefont {Silva}}, \bibinfo {author} {\bibfnamefont {L.~A.}\ \bibnamefont
  {Coelho}}, \bibinfo {author} {\bibfnamefont {F.~W.}\ \bibnamefont {Tavares}},
  \ and\ \bibinfo {author} {\bibfnamefont {M.~J. E.~M.}\ \bibnamefont
  {Cardoso}},\ }\bibfield  {title} {\enquote {\bibinfo {title} {Shear viscosity
  calculated by perturbation theory and molecular dynamics for dense fluids},}\
  }\href@noop {} {\bibfield  {journal} {\bibinfo  {journal} {Inter. J. Quant.
  Chem.}\ }\textbf {\bibinfo {volume} {95}},\ \bibinfo {pages} {79--87}
  (\bibinfo {year} {2003})}\BibitemShut {NoStop}%
\bibitem [{\citenamefont {Heyes}\ and\ \citenamefont
  {Sigurgeirsson}(2004)}]{heyes3}%
  \BibitemOpen
  \bibfield  {author} {\bibinfo {author} {\bibfnamefont {D.~M.}\ \bibnamefont
  {Heyes}}\ and\ \bibinfo {author} {\bibfnamefont {H.}~\bibnamefont
  {Sigurgeirsson}},\ }\bibfield  {title} {\enquote {\bibinfo {title} {The
  {N}ewtonian viscosity of concentrated stabilized dispersions: Comparisons
  with the hard sphere fluid},}\ }\href@noop {} {\bibfield  {journal} {\bibinfo
   {journal} {J. Rheol.}\ }\textbf {\bibinfo {volume} {48}},\ \bibinfo {pages}
  {223} (\bibinfo {year} {2004})}\BibitemShut {NoStop}%
\bibitem [{\citenamefont {van Beijeren}\ and\ \citenamefont
  {Ernst}(1973)}]{Beijeren2}%
  \BibitemOpen
  \bibfield  {author} {\bibinfo {author} {\bibfnamefont {H.}~\bibnamefont {van
  Beijeren}}\ and\ \bibinfo {author} {\bibfnamefont {M.~H.}\ \bibnamefont
  {Ernst}},\ }\bibfield  {title} {\enquote {\bibinfo {title} {The modified
  enskog equation},}\ }\href@noop {} {\bibfield  {journal} {\bibinfo  {journal}
  {Physica}\ }\textbf {\bibinfo {volume} {68}},\ \bibinfo {pages} {437}
  (\bibinfo {year} {1973})}\BibitemShut {NoStop}%
\bibitem [{\citenamefont {de~Haro}\ and\ \citenamefont
  {Cohen}(1984)}]{lopezharo84}%
  \BibitemOpen
  \bibfield  {author} {\bibinfo {author} {\bibfnamefont {M.~López}\
  \bibnamefont {de~Haro}}\ and\ \bibinfo {author} {\bibfnamefont {E.~G.~D.}\
  \bibnamefont {Cohen}},\ }\bibfield  {title} {\enquote {\bibinfo {title} {The
  enskog theory for multicomponent mixtures. iii. transport properties of dense
  binary mixtures with one tracer component},}\ }\href@noop {} {\bibfield
  {journal} {\bibinfo  {journal} {J. Chem. Phys}\ }\textbf {\bibinfo {volume}
  {80}},\ \bibinfo {pages} {408–415} (\bibinfo {year} {1984})}\BibitemShut
  {NoStop}%
\bibitem [{\citenamefont {Kirkpatrick}\ \emph {et~al.}(1990)\citenamefont
  {Kirkpatrick}, \citenamefont {Das}, \citenamefont {Ernst},\ and\
  \citenamefont {Piasecki}}]{Kirkpatrick}%
  \BibitemOpen
  \bibfield  {author} {\bibinfo {author} {\bibfnamefont {T.~R.}\ \bibnamefont
  {Kirkpatrick}}, \bibinfo {author} {\bibfnamefont {Shankar~P.}\ \bibnamefont
  {Das}}, \bibinfo {author} {\bibfnamefont {M.~H.}\ \bibnamefont {Ernst}}, \
  and\ \bibinfo {author} {\bibfnamefont {J.}~\bibnamefont {Piasecki}},\
  }\bibfield  {title} {\enquote {\bibinfo {title} {Kinetic theory of transport
  in a hard sphere crystal},}\ }\href@noop {} {\bibfield  {journal} {\bibinfo
  {journal} {J. Chem. Phys.}\ }\textbf {\bibinfo {volume} {92}},\ \bibinfo
  {pages} {3768–3780} (\bibinfo {year} {1990})}\BibitemShut {NoStop}%
\bibitem [{\citenamefont {Pieprzyk}\ \emph {et~al.}(2024)\citenamefont
  {Pieprzyk}, \citenamefont {Brańka}, \citenamefont {Heyes},\ and\
  \citenamefont {Bannerman}}]{Pieprzyk2024}%
  \BibitemOpen
  \bibfield  {author} {\bibinfo {author} {\bibfnamefont {S.}~\bibnamefont
  {Pieprzyk}}, \bibinfo {author} {\bibfnamefont {A.~C.}\ \bibnamefont
  {Brańka}}, \bibinfo {author} {\bibfnamefont {D.~M.}\ \bibnamefont {Heyes}},
  \ and\ \bibinfo {author} {\bibfnamefont {M.~N.}\ \bibnamefont {Bannerman}},\
  }\bibfield  {title} {\enquote {\bibinfo {title} {Revised enskog theory and
  molecular dynamics simulations of the viscosities and thermal conductivity of
  the hard-sphere fluid and crystal},}\ }\href@noop {} {\bibfield  {journal}
  {\bibinfo  {journal} {Physical Review E}\ }\textbf {\bibinfo {volume}
  {109}},\ \bibinfo {pages} {054119} (\bibinfo {year} {2024})}\BibitemShut
  {NoStop}%
\bibitem [{\citenamefont {Weeks}\ \emph {et~al.}(1971)\citenamefont {Weeks},
  \citenamefont {Chandler},\ and\ \citenamefont {Andersen}}]{weeks}%
  \BibitemOpen
  \bibfield  {author} {\bibinfo {author} {\bibfnamefont {J.D.}\ \bibnamefont
  {Weeks}}, \bibinfo {author} {\bibfnamefont {D.}~\bibnamefont {Chandler}}, \
  and\ \bibinfo {author} {\bibfnamefont {H.C.}\ \bibnamefont {Andersen}},\
  }\bibfield  {title} {\enquote {\bibinfo {title} {The role of repulsive forces
  in determining the equilibrium structure of simple liquids},}\ }\href@noop {}
  {\bibfield  {journal} {\bibinfo  {journal} {J. Chem. Phys.}\ }\textbf
  {\bibinfo {volume} {54}},\ \bibinfo {pages} {5237} (\bibinfo {year}
  {1971})}\BibitemShut {NoStop}%
\bibitem [{\citenamefont {Hildebrand}(1971)}]{hildebrand}%
  \BibitemOpen
  \bibfield  {author} {\bibinfo {author} {\bibfnamefont {J.~H.}\ \bibnamefont
  {Hildebrand}},\ }\bibfield  {title} {\enquote {\bibinfo {title} {Motions of
  molecules in liquids: Viscosity and diffusivity},}\ }\href@noop {} {\bibfield
   {journal} {\bibinfo  {journal} {Science}\ }\textbf {\bibinfo {volume}
  {174}},\ \bibinfo {pages} {490} (\bibinfo {year} {1971})}\BibitemShut
  {NoStop}%
\bibitem [{\citenamefont {Batschinski}(1913)}]{batschinski}%
  \BibitemOpen
  \bibfield  {author} {\bibinfo {author} {\bibfnamefont {A.~J.}\ \bibnamefont
  {Batschinski}},\ }\bibfield  {title} {\enquote {\bibinfo {title}
  {Investigations of internal friction of fluids},}\ }\href@noop {} {\bibfield
  {journal} {\bibinfo  {journal} {Z. Phys. Chem.}\ }\textbf {\bibinfo {volume}
  {84}},\ \bibinfo {pages} {643} (\bibinfo {year} {1913})}\BibitemShut
  {NoStop}%
\bibitem [{\citenamefont {Doolittle}(1951)}]{doolittle}%
  \BibitemOpen
  \bibfield  {author} {\bibinfo {author} {\bibfnamefont {A.~K.}\ \bibnamefont
  {Doolittle}},\ }\bibfield  {title} {\enquote {\bibinfo {title} {Studies in
  {N}ewtonian flow. {II}. {T}he dependence of the viscosity of liquids on
  free‐space},}\ }\href@noop {} {\bibfield  {journal} {\bibinfo  {journal}
  {J. Appl. Phys.}\ }\textbf {\bibinfo {volume} {22}},\ \bibinfo {pages} {1471}
  (\bibinfo {year} {1951})}\BibitemShut {NoStop}%
\bibitem [{\citenamefont {Cohen}\ and\ \citenamefont
  {Turnbull}(1959)}]{cohen2}%
  \BibitemOpen
  \bibfield  {author} {\bibinfo {author} {\bibfnamefont {M.~H.}\ \bibnamefont
  {Cohen}}\ and\ \bibinfo {author} {\bibfnamefont {D.}~\bibnamefont
  {Turnbull}},\ }\bibfield  {title} {\enquote {\bibinfo {title} {Molecular
  transport in liquids and glasses},}\ }\href@noop {} {\bibfield  {journal}
  {\bibinfo  {journal} {J. Chem. Phys.}\ }\textbf {\bibinfo {volume} {31}},\
  \bibinfo {pages} {1164} (\bibinfo {year} {1959})}\BibitemShut {NoStop}%
\bibitem [{\citenamefont {Turnbull}\ and\ \citenamefont
  {Cohen}(1970)}]{turnbull}%
  \BibitemOpen
  \bibfield  {author} {\bibinfo {author} {\bibfnamefont {D.}~\bibnamefont
  {Turnbull}}\ and\ \bibinfo {author} {\bibfnamefont {M.~H.}\ \bibnamefont
  {Cohen}},\ }\bibfield  {title} {\enquote {\bibinfo {title} {On the
  free‐volume model of the liquid‐glass transition},}\ }\href@noop {}
  {\bibfield  {journal} {\bibinfo  {journal} {J. Chem. Phys.}\ }\textbf
  {\bibinfo {volume} {52}},\ \bibinfo {pages} {3038} (\bibinfo {year}
  {1970})}\BibitemShut {NoStop}%
\bibitem [{\citenamefont {Macedo}\ and\ \citenamefont
  {Litovitz}(1965)}]{macedo}%
  \BibitemOpen
  \bibfield  {author} {\bibinfo {author} {\bibfnamefont {P.~B.}\ \bibnamefont
  {Macedo}}\ and\ \bibinfo {author} {\bibfnamefont {T.~A.}\ \bibnamefont
  {Litovitz}},\ }\bibfield  {title} {\enquote {\bibinfo {title} {On the
  relative roles of free volume and activation energy in the viscosity of
  liquids},}\ }\href@noop {} {\bibfield  {journal} {\bibinfo  {journal} {J.
  Chem. Phys.}\ }\textbf {\bibinfo {volume} {42}},\ \bibinfo {pages} {245}
  (\bibinfo {year} {1965})}\BibitemShut {NoStop}%
\bibitem [{\citenamefont {Rosenfeld}(1977{\natexlab{a}})}]{rosenfeld}%
  \BibitemOpen
  \bibfield  {author} {\bibinfo {author} {\bibfnamefont {Y.}~\bibnamefont
  {Rosenfeld}},\ }\bibfield  {title} {\enquote {\bibinfo {title} {Relation
  between the transport coefficients and the internal entropy of simple
  systems},}\ }\href@noop {} {\bibfield  {journal} {\bibinfo  {journal} {Phys.
  Rev. A}\ }\textbf {\bibinfo {volume} {15}},\ \bibinfo {pages} {2545}
  (\bibinfo {year} {1977}{\natexlab{a}})}\BibitemShut {NoStop}%
\bibitem [{\citenamefont {Rosenfeld}(1977{\natexlab{b}})}]{rosenfeld2}%
  \BibitemOpen
  \bibfield  {author} {\bibinfo {author} {\bibfnamefont {Y.}~\bibnamefont
  {Rosenfeld}},\ }\bibfield  {title} {\enquote {\bibinfo {title} {Comments on
  the transport coefficients of dense hard core systems},}\ }\href@noop {}
  {\bibfield  {journal} {\bibinfo  {journal} {Chem. Phys. Lett.}\ }\textbf
  {\bibinfo {volume} {48}},\ \bibinfo {pages} {467} (\bibinfo {year}
  {1977}{\natexlab{b}})}\BibitemShut {NoStop}%
\bibitem [{\citenamefont {Dyre}(2018)}]{dyre}%
  \BibitemOpen
  \bibfield  {author} {\bibinfo {author} {\bibfnamefont {Jeppe~C.}\
  \bibnamefont {Dyre}},\ }\bibfield  {title} {\enquote {\bibinfo {title}
  {Perspective: Excess-entropy scaling},}\ }\href@noop {} {\bibfield  {journal}
  {\bibinfo  {journal} {J. Chem. Phys.}\ }\textbf {\bibinfo {volume} {149}},\
  \bibinfo {pages} {210901} (\bibinfo {year} {2018})}\BibitemShut {NoStop}%
\bibitem [{\citenamefont {Marchioni}\ \emph {et~al.}(2023)\citenamefont
  {Marchioni}, \citenamefont {Muro},\ and\ \citenamefont
  {Hoyuelos}}]{marchioni}%
  \BibitemOpen
  \bibfield  {author} {\bibinfo {author} {\bibfnamefont {L.}~\bibnamefont
  {Marchioni}}, \bibinfo {author} {\bibfnamefont {M.~A.~Di}\ \bibnamefont
  {Muro}}, \ and\ \bibinfo {author} {\bibfnamefont {M.}~\bibnamefont
  {Hoyuelos}},\ }\bibfield  {title} {\enquote {\bibinfo {title} {Dependence on
  the thermodynamic state of self-diffusion of pseudo-hard-sphere and
  lennard-jones potentials},}\ }\href@noop {} {\bibfield  {journal} {\bibinfo
  {journal} {Phys. Rev. E}\ }\textbf {\bibinfo {volume} {107}},\ \bibinfo
  {pages} {014134} (\bibinfo {year} {2023})}\BibitemShut {NoStop}%
\bibitem [{\citenamefont {Cohen}\ and\ \citenamefont
  {de~Schepper}(1991)}]{cohen3}%
  \BibitemOpen
  \bibfield  {author} {\bibinfo {author} {\bibfnamefont {E.~G.~D.}\
  \bibnamefont {Cohen}}\ and\ \bibinfo {author} {\bibfnamefont {I.~M.}\
  \bibnamefont {de~Schepper}},\ }\bibfield  {title} {\enquote {\bibinfo {title}
  {Note on transport processes in dense colloidal suspensions},}\ }\href@noop
  {} {\bibfield  {journal} {\bibinfo  {journal} {Journal of Statistical
  Physics}\ }\textbf {\bibinfo {volume} {63}},\ \bibinfo {pages} {241}
  (\bibinfo {year} {1991})}\BibitemShut {NoStop}%
\bibitem [{\citenamefont {Plimpton}(1995)}]{plimpton}%
  \BibitemOpen
  \bibfield  {author} {\bibinfo {author} {\bibfnamefont {S.}~\bibnamefont
  {Plimpton}},\ }\bibfield  {title} {\enquote {\bibinfo {title} {Fast parallel
  algorithms for short-range molecular dynamics},}\ }\href@noop {} {\bibfield
  {journal} {\bibinfo  {journal} {J. Comput. Phys.}\ }\textbf {\bibinfo
  {volume} {117}},\ \bibinfo {pages} {1--19} (\bibinfo {year}
  {1995})}\BibitemShut {NoStop}%
\bibitem [{\citenamefont {{Di Muro}}\ and\ \citenamefont
  {Hoyuelos}(2021)}]{dimuro}%
  \BibitemOpen
  \bibfield  {author} {\bibinfo {author} {\bibfnamefont {M.~A.}\ \bibnamefont
  {{Di Muro}}}\ and\ \bibinfo {author} {\bibfnamefont {M.}~\bibnamefont
  {Hoyuelos}},\ }\bibfield  {title} {\enquote {\bibinfo {title} {Application of
  the {W}idom insertion formula to transition rates in a lattice},}\
  }\href@noop {} {\bibfield  {journal} {\bibinfo  {journal} {Phys. Rev. E}\
  }\textbf {\bibinfo {volume} {104}},\ \bibinfo {pages} {044104} (\bibinfo
  {year} {2021})}\BibitemShut {NoStop}%
\bibitem [{\citenamefont {Pathria}\ and\ \citenamefont
  {Beale}(2011)}]{pathria}%
  \BibitemOpen
  \bibfield  {author} {\bibinfo {author} {\bibfnamefont {R.~K.}\ \bibnamefont
  {Pathria}}\ and\ \bibinfo {author} {\bibfnamefont {P.~D.}\ \bibnamefont
  {Beale}},\ }\href@noop {} {\emph {\bibinfo {title} {Statistical
  Mechanics}}},\ \bibinfo {edition} {3rd}\ ed.\ (\bibinfo  {publisher}
  {Elsevier},\ \bibinfo {year} {2011})\BibitemShut {NoStop}%
\bibitem [{\citenamefont {Hirschfelder}\ \emph {et~al.}(1954)\citenamefont
  {Hirschfelder}, \citenamefont {Curtiss},\ and\ \citenamefont
  {Bird}}]{Hirschfelder}%
  \BibitemOpen
  \bibfield  {author} {\bibinfo {author} {\bibfnamefont {J.~O.}\ \bibnamefont
  {Hirschfelder}}, \bibinfo {author} {\bibfnamefont {C.~F.}\ \bibnamefont
  {Curtiss}}, \ and\ \bibinfo {author} {\bibfnamefont {R.~B.}\ \bibnamefont
  {Bird}},\ }\href@noop {} {\emph {\bibinfo {title} {Molecular Theory of Gases
  and Liquids}}}\ (\bibinfo  {publisher} {Wiley},\ \bibinfo {address} {New
  York},\ \bibinfo {year} {1954})\BibitemShut {NoStop}%
\bibitem [{\citenamefont {Fokin}\ \emph {et~al.}(1999)\citenamefont {Fokin},
  \citenamefont {Popov},\ and\ \citenamefont {Kalashnikov}}]{fokin}%
  \BibitemOpen
  \bibfield  {author} {\bibinfo {author} {\bibfnamefont {L.~R.}\ \bibnamefont
  {Fokin}}, \bibinfo {author} {\bibfnamefont {V.N.}\ \bibnamefont {Popov}}, \
  and\ \bibinfo {author} {\bibfnamefont {A.N.}\ \bibnamefont {Kalashnikov}},\
  }\bibfield  {title} {\enquote {\bibinfo {title} {Analytical representation of
  collision integrals for the (m-6) {L}ennard-{J}ones potentials in the
  {EPIDIF} database},}\ }\href@noop {} {\bibfield  {journal} {\bibinfo
  {journal} {High Temp.}\ }\textbf {\bibinfo {volume} {37}},\ \bibinfo {pages}
  {45} (\bibinfo {year} {1999})}\BibitemShut {NoStop}%
\bibitem [{\citenamefont {McQuarrie}(2000)}]{mcquarrie}%
  \BibitemOpen
  \bibfield  {author} {\bibinfo {author} {\bibfnamefont {D.~A.}\ \bibnamefont
  {McQuarrie}},\ }\href@noop {} {\emph {\bibinfo {title} {Statistical
  Mechanics}}}\ (\bibinfo  {publisher} {University Science Books},\ \bibinfo
  {address} {Sausalito},\ \bibinfo {year} {2000})\BibitemShut {NoStop}%
\bibitem [{\citenamefont {Reichl}(1998)}]{reichl}%
  \BibitemOpen
  \bibfield  {author} {\bibinfo {author} {\bibfnamefont {L.~E.}\ \bibnamefont
  {Reichl}},\ }\href@noop {} {\emph {\bibinfo {title} {A Modem Course in
  Statistical Physics}}},\ \bibinfo {edition} {2nd}\ ed.\ (\bibinfo
  {publisher} {Wiley},\ \bibinfo {year} {1998})\BibitemShut {NoStop}%
\bibitem [{\citenamefont {Jover}\ \emph {et~al.}(2012)\citenamefont {Jover},
  \citenamefont {Haslam}, \citenamefont {Galindo}, \citenamefont {Jackson},\
  and\ \citenamefont {Müller}}]{jover}%
  \BibitemOpen
  \bibfield  {author} {\bibinfo {author} {\bibfnamefont {J.}~\bibnamefont
  {Jover}}, \bibinfo {author} {\bibfnamefont {A.~J.}\ \bibnamefont {Haslam}},
  \bibinfo {author} {\bibfnamefont {A.}~\bibnamefont {Galindo}}, \bibinfo
  {author} {\bibfnamefont {G.}~\bibnamefont {Jackson}}, \ and\ \bibinfo
  {author} {\bibfnamefont {E.~A.}\ \bibnamefont {Müller}},\ }\bibfield
  {title} {\enquote {\bibinfo {title} {Pseudo hard-sphere potential for use in
  continuous molecular-dynamics simulation of spherical and chain molecules},}\
  }\href@noop {} {\bibfield  {journal} {\bibinfo  {journal} {J. Chem. Phys.}\
  }\textbf {\bibinfo {volume} {137}},\ \bibinfo {pages} {144505} (\bibinfo
  {year} {2012})}\BibitemShut {NoStop}%
\bibitem [{\citenamefont {Pousaneh}\ and\ \citenamefont
  {de~Wijn}(2019)}]{Pousaneh}%
  \BibitemOpen
  \bibfield  {author} {\bibinfo {author} {\bibfnamefont {Faezeh}\ \bibnamefont
  {Pousaneh}}\ and\ \bibinfo {author} {\bibfnamefont {Astrid~S.}\ \bibnamefont
  {de~Wijn}},\ }\bibfield  {title} {\enquote {\bibinfo {title} {Shear viscosity
  of pseudo hard-spheres},}\ }\href@noop {} {\bibfield  {journal} {\bibinfo
  {journal} {Molecular Physics}\ }\textbf {\bibinfo {volume} {118}} (\bibinfo
  {year} {2019})}\BibitemShut {NoStop}%
\bibitem [{\citenamefont {Pieprzyk}\ \emph {et~al.}(2019)\citenamefont
  {Pieprzyk}, \citenamefont {Bannerman}, \citenamefont {Bra\'{n}ka},
  \citenamefont {Chudak},\ and\ \citenamefont {Heyes}}]{pieprzyk2}%
  \BibitemOpen
  \bibfield  {author} {\bibinfo {author} {\bibfnamefont {S.}~\bibnamefont
  {Pieprzyk}}, \bibinfo {author} {\bibfnamefont {M.~N.}\ \bibnamefont
  {Bannerman}}, \bibinfo {author} {\bibfnamefont {A.~C.}\ \bibnamefont
  {Bra\'{n}ka}}, \bibinfo {author} {\bibfnamefont {M.}~\bibnamefont {Chudak}},
  \ and\ \bibinfo {author} {\bibfnamefont {D.~M.}\ \bibnamefont {Heyes}},\
  }\bibfield  {title} {\enquote {\bibinfo {title} {Thermodynamic and dynamical
  properties of the hard sphere system revisited by molecular dynamics
  simulation},}\ }\href@noop {} {\bibfield  {journal} {\bibinfo  {journal}
  {Phys. Chem. Chem. Phys.}\ }\textbf {\bibinfo {volume} {21}},\ \bibinfo
  {pages} {6886} (\bibinfo {year} {2019})}\BibitemShut {NoStop}%
\bibitem [{\citenamefont {Mulero}\ \emph {et~al.}(2008)\citenamefont {Mulero},
  \citenamefont {Galán}, \citenamefont {Parra},\ and\ \citenamefont
  {Cuadros}}]{mulero1}%
  \BibitemOpen
  \bibfield  {author} {\bibinfo {author} {\bibfnamefont {A.}~\bibnamefont
  {Mulero}}, \bibinfo {author} {\bibfnamefont {C.~A.}\ \bibnamefont {Galán}},
  \bibinfo {author} {\bibfnamefont {M.~I.}\ \bibnamefont {Parra}}, \ and\
  \bibinfo {author} {\bibfnamefont {F.}~\bibnamefont {Cuadros}},\ }\bibfield
  {title} {\enquote {\bibinfo {title} {Equations of state for hard spheres and
  hard disks},}\ }in\ \href@noop {} {\emph {\bibinfo {booktitle} {Theory and
  Simulation of Hard-Sphere Fluids and Related Systems}}},\ \bibinfo {editor}
  {edited by\ \bibinfo {editor} {\bibfnamefont {A.}~\bibnamefont {Mulero}}}\
  (\bibinfo  {publisher} {Springer},\ \bibinfo {year} {2008})\ p.~\bibinfo
  {pages} {37}\BibitemShut {NoStop}%
\bibitem [{\citenamefont {Meier}\ \emph {et~al.}(2004)\citenamefont {Meier},
  \citenamefont {Laesecke},\ and\ \citenamefont {Kabelac}}]{meier3}%
  \BibitemOpen
  \bibfield  {author} {\bibinfo {author} {\bibfnamefont {K.}~\bibnamefont
  {Meier}}, \bibinfo {author} {\bibfnamefont {A.}~\bibnamefont {Laesecke}}, \
  and\ \bibinfo {author} {\bibfnamefont {S.}~\bibnamefont {Kabelac}},\
  }\bibfield  {title} {\enquote {\bibinfo {title} {Transport coefficients of
  the lennard-jones model fluid. i. viscosity},}\ }\href@noop {} {\bibfield
  {journal} {\bibinfo  {journal} {J. Chem. Phys.}\ }\textbf {\bibinfo {volume}
  {121}},\ \bibinfo {pages} {3671–3687} (\bibinfo {year} {2004})}\BibitemShut
  {NoStop}%
\bibitem [{\citenamefont {Galli\'ero}\ and\ \citenamefont
  {Baylaucq}(2005)}]{Galliero}%
  \BibitemOpen
  \bibfield  {author} {\bibinfo {author} {\bibfnamefont {G.}~\bibnamefont
  {Galli\'ero}}\ and\ \bibinfo {author} {\bibfnamefont {C.~Bonedand~A.}\
  \bibnamefont {Baylaucq}},\ }\bibfield  {title} {\enquote {\bibinfo {title}
  {Molecular dynamics study of the lennard-jones fluid viscosity: Application
  to real fluids},}\ }\href@noop {} {\bibfield  {journal} {\bibinfo  {journal}
  {Ind. Eng. Chem. Res.}\ }\textbf {\bibinfo {volume} {44}},\ \bibinfo {pages}
  {6963--6972} (\bibinfo {year} {2005})}\BibitemShut {NoStop}%
\bibitem [{\citenamefont {Bird}\ \emph {et~al.}(2015)\citenamefont {Bird},
  \citenamefont {Stewart}, \citenamefont {Lightfoot},\ and\ \citenamefont
  {Klingenberg}}]{Bird}%
  \BibitemOpen
  \bibfield  {author} {\bibinfo {author} {\bibfnamefont {R.~Byron}\
  \bibnamefont {Bird}}, \bibinfo {author} {\bibfnamefont {Warren~E.}\
  \bibnamefont {Stewart}}, \bibinfo {author} {\bibfnamefont {Edwin~N.}\
  \bibnamefont {Lightfoot}}, \ and\ \bibinfo {author} {\bibfnamefont
  {Daniel~J.}\ \bibnamefont {Klingenberg}},\ }\href@noop {} {\emph {\bibinfo
  {title} {Introductory Transport Phenomena, 1st Edition}}},\ \bibinfo
  {edition} {1st}\ ed.\ (\bibinfo  {publisher} {Wiley},\ \bibinfo {year}
  {2015})\BibitemShut {NoStop}%
\end{thebibliography}%

\end{document}